\documentclass[twocolumn]{aastex701}
\usepackage{courier}
\usepackage{float}
\usepackage{tabularx}
\usepackage{amsmath}
\usepackage[figuresright]{rotating}
\usepackage{amssymb}
\usepackage{graphicx}
\usepackage{natbib}

\newcolumntype{L}{>{\raggedright\arraybackslash}p{1.4cm}}
\newcolumntype{M}{>{\centering\arraybackslash}p{2.2cm}}
\newcolumntype{D}{>{\raggedright\arraybackslash}X}
\newcolumntype{O}{>{\raggedright\arraybackslash}p{1.9cm}}
\newcolumntype{R}{>{\raggedleft\arraybackslash}p{2.5cm}}

\newcommand{\Msol}{\hbox{M$_{\sun}$}}

\newcommand{\eg}{e.g.}

\newcommand{\etal}{\hbox{et~al.}}

\newcommand\arcspt       {{$\buildrel{\prime\prime}\over .$}}

\defcitealias{Zhao2024}{Z24}
\defcitealias{Silver2026}{S26}
\newcommand{\Zhao}{\citetalias{Zhao2024}}
\newcommand{\Silver}{\citetalias{Silver2026}}

\begin{document}

\title{PEARLS: NuSTAR and XMM-Newton Extragalactic Survey of the JWST North Ecliptic Pole
Time-Domain Field VI: Multiwavelength SED Analysis}
\shortauthors{Ortiz \etal}
\shorttitle{X-ray AGN and Their Host Galaxies in the NEP TDF}

\author[0000-0002-6150-833X]{Rafael Ortiz~III}
\affiliation{School of Earth and Space Exploration, Arizona State University,
Tempe, AZ 85287-1404, USA}
\email{rortizii@asu.edu}

\author[0000-0002-2115-1137]{Francesca Civano}
\affiliation{NASA Goddard Space Flight Center, Greenbelt, MD 20771, USA}
\email{francesca.m.civano@nasa.gov}

\author[0000-0001-8156-6281]{Rogier A. Windhorst}
\affiliation{School of Earth and Space Exploration, Arizona State University, Tempe, AZ 85287-1404, USA}
\email{Rogier.Windhorst@gmail.com}

\author[0000-0002-9895-5758]{S.\ P.\ Willner}
\affiliation{Center for Astrophysics \textbar\ Harvard \& Smithsonian, 60 Garden Street, Cambridge, MA, 02138, USA}
\email{swillner@cfa.harvard.edu}


\author[0009-0007-0782-0721]{Gibson B. Bowling}
\affiliation{School of Earth and Space Exploration, Arizona State University, Tempe, AZ 85287-1404, USA}
\email{gbbowlin@asu.edu}

\author[0000-0001-6650-2853]{Timothy Carleton}
\email{tmcarlet@asu.edu}
\affiliation{School of Earth and Space Exploration, Arizona State University,
Tempe, AZ 85287-1404, USA}

\author[0000-0003-3329-1337]{Seth H. Cohen}
\affiliation{School of Earth and Space Exploration, Arizona State University, Tempe, AZ 85287-1404, USA}
\email{seth.cohen@asu.edu}

\author[0000-0002-9041-7437]{Samantha Creech}
\affiliation{Department of Physics and Astronomy, University of Utah, 115 South 1400 East, Salt Lake City, UT 84112, USA}
\email{s.creech@utah.edu}

\author[0000-0001-8489-2349]{Vicente Estrada-Carpenter}
\affiliation{School of Earth and Space Exploration, Arizona State University, Tempe, AZ 85287-1404, USA}
\email{vestrad9@asu.edu}

\author[0000-0003-1625-8009]{Brenda L.~Frye}
\email{brendafrye@gmail.com}
\affiliation{Department of Astronomy/Steward Observatory, University of Arizona, 933 N Cherry Ave,
Tucson, AZ, 85721-0009, USA}

\author[0000-0001-9440-8872]{Norman A.~Grogin}
\affiliation{Space Telescope Science Institute, 3700 San Martin Drive,
Baltimore, MD 21218, USA}
\email{nagrogin@stsci.edu}

\author[0000-0001-8751-3463]{Heidi B.~Hammel}
\email{hbhammel@aura-astronomy.org}
\affiliation{Association of Universities for Research in Astronomy, 1331 Pennsylvania 
Avenue NW, Suite 1475, Washington, DC 20005, USA}

\author[0000-0001-6670-6370]{Timothy Heckman}
\affiliation{William Miller III Department of Physics \& Astronomy, The Johns Hopkins University, Baltimore, MD 21218}
\affiliation{School of Earth and Space Exploration, Arizona State University, Tempe, AZ 85287-1404, USA}
\email{theckma1@jhu.edu}

\author[0000-0002-9984-4937]{Rachel Honor}
\email{rchonor@asu.edu}
\affiliation{School of Earth and Space Exploration, Arizona State University, Tempe, AZ 85287-1404, USA}

\author[0000-0003-1268-5230]{Rolf A. Jansen}
\affiliation{School of Earth and Space Exploration, Arizona State University, Tempe, AZ 85287-1404, USA}
\email{rolfjansen.work@gmail.com}

\author[0000-0003-3214-9128]{Satoshi Kikuta}
\affiliation{National Astronomical Observatory of Japan, 2-21-1 Osawa, Mitaka, Tokyo 181-8588, Japan,}
\email{kikuta.astro@gmail.com}

\author[0000-0002-6610-2048]{Anton M. Koekemoer}
\affiliation{Space Telescope Science Institute, 3700 San Martin Drive, Baltimore, MD 21218, USA}
\email{koekemoer@stsci.edu}

\author[0000-0001-6434-7845]{Madeline A. Marshall}
\affiliation{Los Alamos National Laboratory, Los Alamos, NM 87545, USA}
\email{madeline_marshall@outlook.com}

\author[0009-0002-3259-1282]{Sylvia Mesicek}
\affiliation{Department of Physics and Astronomy, University of Utah, 115 South 1400 East, Salt Lake City, UT 84112, USA}
\email{sylvia.mesicek@utah.edu}

\author[0000-0003-4440-259X]{Mar Mezcua}
\affiliation{Institute of Space Sciences (ICE, CSIC), Campus UAB, Carrer de Can Magrans, s/n, 08193 Barcelona, Spain}
\affiliation{Institut d'Estudis Espacials de Catalunya (IEEC),  Edifici RDIT, Campus UPC, 08860 Castelldefels, Barcelona, Spain}
\email{marmezcua.astro@gmail.com}

\author[0000-0001-7694-4129]{Stefanie N.~Milam}
\email{stefanie.n.milam@nasa.gov}
\affiliation{NASA Goddard Space Flight Center, Greenbelt, MD 20771, USA}

\author[0000-0002-5573-9131]{Simon D. Mork}
\affiliation{School of Earth and Space Exploration, Arizona State University, Tempe, AZ 85287-1404, USA}
\email{sdmork@asu.edu}

\author[0000-0003-3351-0878]{Rosalia O'Brien}
\affiliation{Department of Astronomy, University of Maryland, College Park, MD 20742, USA}
\affiliation{Astrophysics Science Division, Code 660, NASA Goddard Space Flight Center, 8800 Greenbelt Rd., Greenbelt, MD 20771, USA}
\affiliation{Center for Research and Exploration in Space Science and Technology, NASA/GSFC, Greenbelt, MD 20771 USA}
\email{rosalia.d.obrien@nasa.gov}

\author[0000-0002-5319-6620]{Payaswini Saikia}
\affiliation{Department of Astronomy, Yale University, PO Box 208101, New Haven, CT 06520-8101, USA}
\email{payaswini.ssc@gmail.com}

\author[0000-0001-6564-0517]{Ross M. Silver}
\affiliation{NASA Goddard Space Flight Center, Greenbelt, MD 20771, USA}
\affiliation{Southeastern Universities Research Association, Washington, DC 20005, USA}
\email{ross.m.silver@nasa.gov}

\author[0000-0002-0648-1699]{Brent M. Smith}
\affiliation{School of Earth and Space Exploration, Arizona State University, Tempe, AZ 85287-1404, USA}
\email{bsmith18@asu.edu}

\author[0000-0002-2536-1633]{Hyewon Suh}
\affiliation{International Gemini Observatory/NSF NOIRLab, 670 N. A'ohoku Place, Hilo, HI 96720, USA}
\email{hyewon.suh@noirlab.edu}

\author[0000-0001-9262-9997]{Christopher N. A. Willmer}
\affiliation{Steward Observatory, University of Arizona, 933 N Cherry Ave, Tucson, AZ, 85721-0009, USA}
\email{cnawillmer@gmail.com}

\author[0000-0001-7592-7714]{Haojing Yan}
\affiliation{Department of Physics and Astronomy, University of Missouri, Columbia, MO 65211, USA}
\email{yanhaojing@gmail.com}

\author[0000-0002-7791-3671]{Xiurui Zhao}
\affiliation{Cahill Center for Astrophysics, California Institute of Technology, 1216 East California Boulevard, Pasadena, CA 91125, USA}
\email{xiurui.zhao.work@gmail.com}


\begin{abstract}

We model spectral energy distributions of 261 X-ray sources to $z \sim 5$ in the North Ecliptic Pole Time Domain Field, extending prior XMM-Newton and NuSTAR analyses. Using the star-forming main sequence (SFMS) and black hole accretion rate (BHAR) frameworks, we find that SFRs generally lie below the SFMS while most BHARs exceed the population average, as expected for X-ray-selected samples. There is a strong correlation ($\rho=+0.73$) between SFR relative to the SFMS and specific AGN luminosity, $L_{\mathrm{AGN}}/M_*$; galaxies with the highest $L_{\mathrm{AGN}}/M_*$ exist at or above the SFMS. X-ray luminosity correlates with SFR ($\rho=+0.80$), revealing a star-forming and X-ray luminous ``cold quasar'' population consistent with dramatic, short-timescale accretion episodes. Low-mass galaxies show BHARs well above the population averaged value for their mass whereas high-mass galaxies' SMBHs accrete at the population averaged BHAR, suggesting ``growth spurt'' and ``maintenance-mode'' accretion, respectively. Traditional AGN classifications (obscured, unobscured, or radio-loud) do not reveal these distinctions, demonstrating the X-ray perspective's unique ability to identify rare AGN phases that are critical for the instantaneous link between galaxies and their SMBHs.

\end{abstract}
\correspondingauthor{Rafael Ortiz III}
\email{rortizii@asu.edu}

\section{Introduction}

Hubble Space Telescope (HST) observations have revealed that supermassive black holes (SMBHs) are nearly ubiquitous in the centers of massive galaxies, and their masses ($M_{\rm BH}$) correlate tightly with the stellar mass ($M_*$) of the galactic spheroid and its velocity dispersion ($\sigma$) \citep[\eg][]{Magorrian_1998,Ferrarese2000, Kormendy2013, Heckman2014}, implying a link between SMBH growth and stellar mass assembly in a galaxy. Further, the similar redshift evolution of the cosmic star-formation rate density (SFRD) and the black hole accretion rate density (BHARD) both peaking at $z\sim2$ suggest that, when averaged over large populations and long timescales, galaxies and their SMBH grow in lockstep \citep{Hopkins2006, Madau2014, Brandt_2015, Aird2015, DSilva2025}.

The X-ray regime is critical towards providing a direct view of radiatively efficient SMBH accretion, since luminous X-ray point sources are secure active galactic nuclei (AGN) tracers with minimal contamination from energetic stellar processes, and because X-rays are sensitive to AGN even with some degree of obscuration \citep{Brandt_2015, Hickox_2018}. However, X-ray data alone cannot contextualize these AGN within their host galaxies because X-ray emission is insensitive to stellar populations, which are best sampled in the visible and near-infrared wavelengths. This motivates multiwavelength analyses of X-ray selected AGN to place radiatively efficient SMBHs within the context of their host galaxies. In particular, spectral energy distribution (SED) fitting has become a common and powerful tool that can provide inferences on $M_*$, star-formation rate (SFR), the star-formation history (SFH), and the AGN emission from the centrally accreting SMBH. 

Central in recent AGN-host studies has been the star-forming main sequence (SFMS), which suggests that galaxies' star-formation rates (SFR) and stellar masses ($M_*$) evolve significantly over cosmic time \citep{Speagle2014, Tomczak2016, Leja2022, Popesso2023}. 
A useful approach is to express a galaxy's SFR as an offset from the star-formation main sequence value, given $M_*$ and redshift. This offset, called $\Delta{\rm SFMS}$, effectively distinguishes between three regimes: quenched galaxies with little to no star-formation, main-sequence star-forming galaxies with typical star-formation, and starburst galaxies with extremely high star-formation.
Studying X-ray selected AGN in different sub-populations based on their SFR offset from the SFMS yields a nuanced picture of the coevolution of AGN and their host galaxies.

\citet{Rosario2013} found that many moderate-luminosity X-ray selected AGN reside in SFMS-like host galaxies, suggesting stochastic fueling in the star-forming galaxy. As X-ray surveys became rich in number counts and ancillary wavelength coverage, well-sampled SEDs of X-ray selected AGN helped to disentangle the relationship between SFR, $M_*$, and X-ray luminosity ($L_X$). \citet{Suh2019} suggested X-ray AGN reside in normal star-forming galaxies and that any link between AGN activity and star-formation is driven by stellar mass rather than strong AGN-driven quenching or boosting of SFR at fixed stellar mass. \citet{Cristello2024} found SFMS offsets in low mass ($\log(M_*/M_\odot)\simeq9.5$--10.5~\Msol) galaxies suggestive of starburst activity, whereas \citet{Mountrichas2022, Mountrichas2023} show that the most X-ray luminous AGN ($L_X\gtrsim10^{44}\,\mathrm{erg\,s^{-1}}$) tend to inhabit galaxies with elevated SFRs, especially for intermediate mass ($10.5 \leq \log\,(M_{\star}/M_{\odot}) \leq 11.5$) galaxies. In addition, \citet{Kirkpatrick2020} finds that some highly luminous quasars host enhanced star-formation, coining ``cold quasars'' as unobscured, luminous, and starburst-AGN systems. These works together show that the average SFR of AGN hosts tracks the SFMS; however, the underlying SFR distribution reflects the dynamic timescales of galaxy evolution and helps situate AGN-galaxy coevolution for the starburst and quenched populations \citep{Mullaney2015, Aird2019}.

Another powerful tool toward understanding the SMBH-galaxy relationship has been the characterization of SMBH accretion rates across diverse host-galaxy demographics. Work by \citet{Aird2018} demonstrates that for star-forming galaxies, the probability of AGN with a given accretion rate is a broad, approximately power-law-like distribution proportional to the Eddington luminosity ($L_{\rm Edd}$). This distribution reflects the duty cycle of AGN as the Eddington ratio ($\lambda_{\rm Edd}$) has been shown to anti-correlate with line-of-sight column density ($N_H$); this radiation-regulated unification scenario presented by \citet{Almeida2017, Ananna2022, Ricci2022} explains radiation driven AGN outflows on gas and dust as the physical mechanism regulating which AGN phase the central SMBH reflects. While the ratio of BH accretion rate (BHAR) to SFR is constant ($\sim10^{-3}$), the instantaneous X-ray luminosity and SFR of individual galaxies are weakly correlated due to the dramatic stochasticities associated with accretion, star-formation, and AGN feedback. Recent work combining multiple X-ray surveys has built a ``population-averaged'' BHAR, $\overline{\rm BHAR}$ \citep{Yang2018, Zou2024}, providing a baseline to which an individual AGN energy output can be compared, thus leveraging the critical vantage point of X-ray emissions for understanding how the dynamic timescales of gas accretion onto the central SMBH contribute to galaxy evolution.

The JWST North Ecliptic Pole (NEP) Time Domain Field (TDF) is a prime field for AGN science due to its deep,  multi-epoch X-ray observations. \citet{Zhao2024} (\Zhao\ hereafter) and \citet{Silver2026} (\Silver\ hereafter) published X-ray catalogs with multiwavelength counterparts in the ultraviolet, visible, and infrared wavelengths. The field has high-resolution, deep space-based imaging from the HST \citep{Obrien2024} and from the James Webb Space Telescope (JWST) \citep{Windhorst2023}. Studies of Compton-thick AGN \citep{Creech2025}, JWST/NIRCam-identified Seyferts with point-like AGN emission \citep{Ortiz2024}, and radio AGN \citep{Willner2026, Saikia2025} are among the analyses to date of unique AGN demographics in the field. Still, the rich multiwavelength datasets have not yet to be combined to fully complement the X-ray sample.

This work compiles the available photometry for all X-ray sources presented in \Zhao\ and \Silver\ to fit their SEDs, providing redshifts and physical inferences for the stellar and AGN emissions. Section~\ref{sec:data} details the multiwavelength datasets incorporated into the analysis, along with the data compilation and catalog creation. Section~\ref{sec:methods} presents the SED fitting,  and Section~\ref{sec:results} synthesizes inferences from these SED fits. Section~\ref{sec:discussion} discusses the results with respect to other multiwavelength X-ray analyses and discusses their implications. Throughout, we adopt a flat Lambda Cold Dark Matter ($\Lambda\rm{CDM}$) cosmology with $H_0 = 67$~{km}~{s$^{-1}$~{Mpc$^{-1}$, $\Omega_\Lambda=0.69$, and $\Omega_{\rm{M}}=0.31$ \citep{Planck2020}.

\section{Data \& Catalogs}
\label{sec:data}

\begin{figure}
    \centering
    \hspace{-0.2cm}
    \includegraphics[width=0.99\linewidth,clip=true,trim= 0 0 0 65]{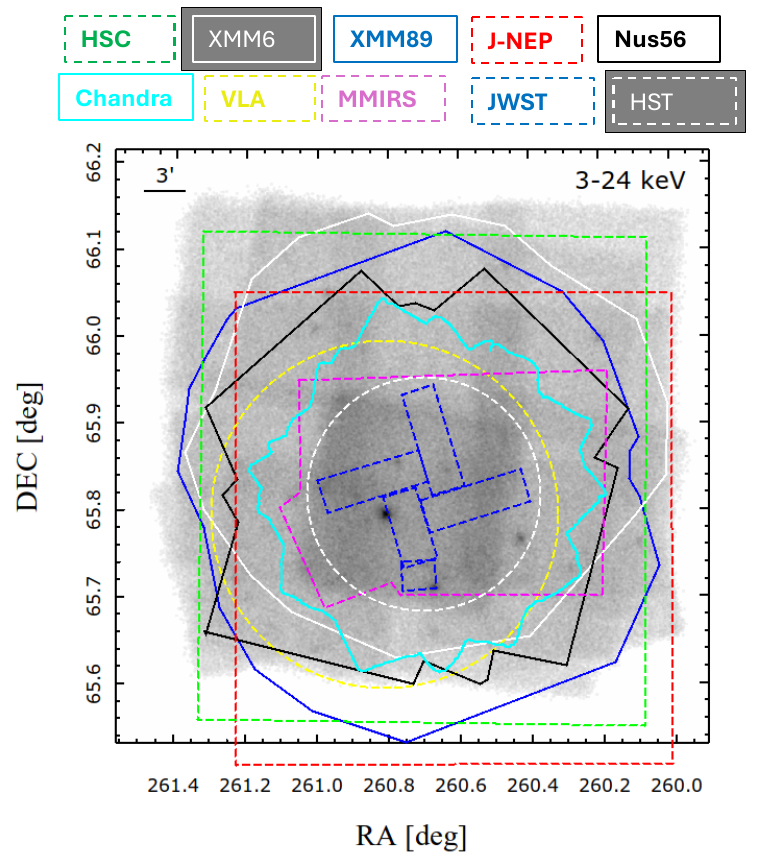}
    \caption{Survey footprints from the various observations in the NEP TDF from \citet[][their Fig.~13]{Silver2026}. The background negative image shows the \hbox{full}-depth NuSTAR image.  The solid white (Cy~6) and solid blue (Cy~8+9) lines shows the XMM area from which X-ray sources were selected, and the dashed green outline shows the HSC coverage that was the basis for counterpart identification.  Other lines show the Chandra (solid cyan), VLA (dashed yellow), MMIRS (dashed magenta), HST (dashed white), and JWST/NIRCam (dashed blue) coverage.}
    \label{fig:NEP}
\end{figure}

\begin{figure}
    \centering
    \hspace{-0.2cm}
    \includegraphics[width=0.99\linewidth]{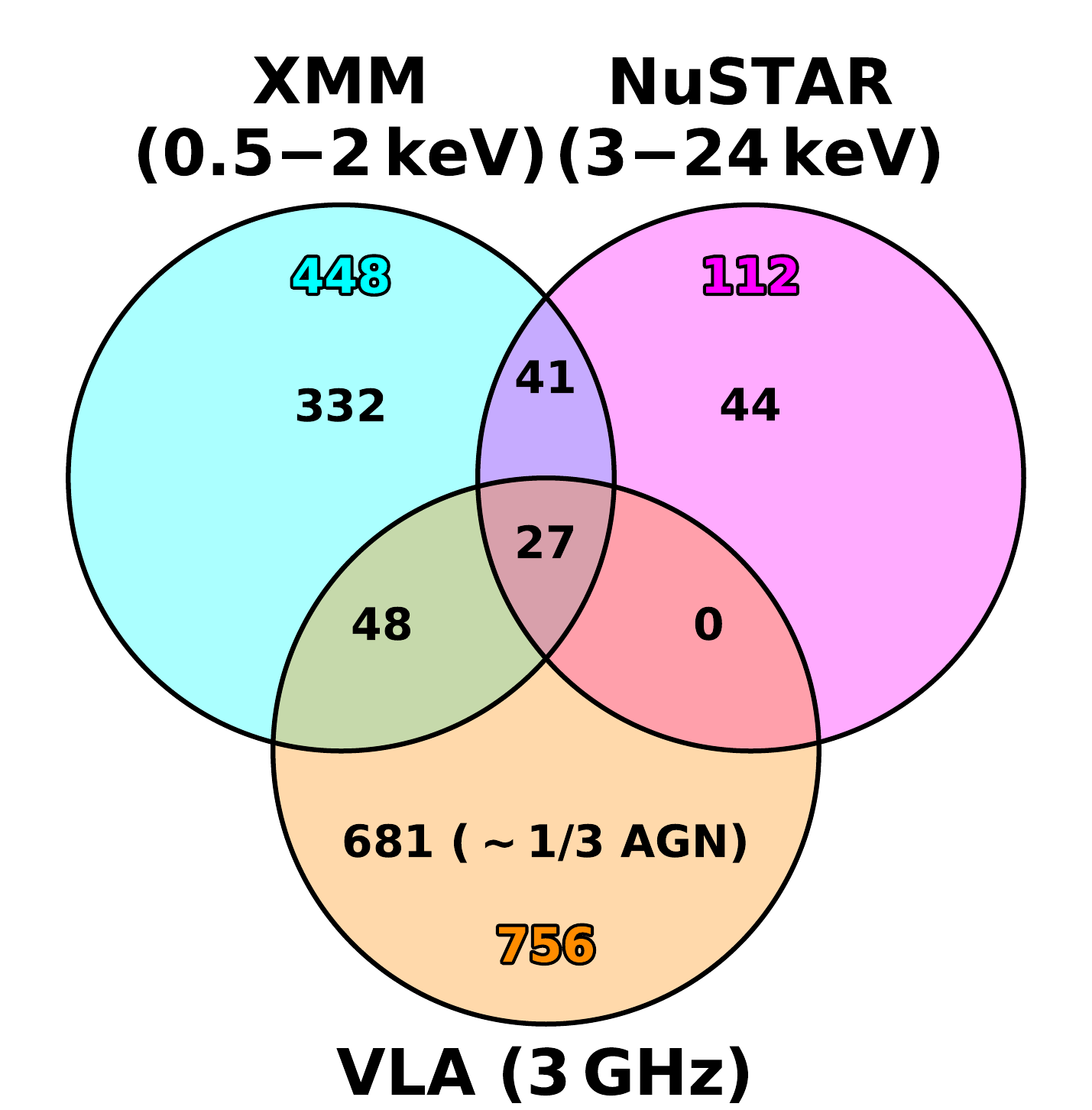}
    \caption{Venn Diagram showing source overlap in the NEP TDF between XMM-Newton, NuSTAR, and VLA from identifications presented in \Zhao , \Silver , and \citet{Hyun2023}. Totals for each are indicated in color and bold outline.}
    \label{fig:venn}
\end{figure}

\subsection{X-ray Data and Sample Selection}

X-ray sources studied in this work are those in the \Zhao\ and \Silver\ catalogs of XMM Cycle 6 and Cycle 8+9 observations, respectively.
The combined catalogs present 453 unique sources.
The X-ray observations of the NEP TDF include those with the Nuclear Spectroscopic Telescope Array (NuSTAR) in addition to those with XMM-Newton to cover the energy range 0.5--24~keV. In total, the observations include 3.5~Ms of cumulative NuSTAR data and 228~ks of XMM data over 0.31~deg$^2$.

The SED analysis requires that X-ray fluxes are input as intrinsic fluxes  corrected for absorption. Absorption-corrected X-ray fluxes were measured from the NuSTAR and XMM-Newton data using the \textsc{xspec} \citep{Arnaud1996} spectral-analysis package.  \citet{Creech2025} and Creech et al.\ (2026 in prep.)\  analyzed NuSTAR-selected catalogs from Cycles 5+6 and 8+9, respectively, and  Mesicek et al.\ (in prep.)\ will report the XMM-Newton analysis. In brief, the X-ray spectra were fit to obscured power-law models \citep[][their Eq.~1]{Creech2025} to measure the intrinsic power-law and ($N_{\mathrm{H}}$)\null. 
Appendix Figure~\ref{fig:intrinsic-flux} shows the differences between observed and intrinsic fluxes for the sample.

While X-ray binary (XRB) emission associated with star-formation can also be present in AGN hosts,  $z \la 1$ sources show upper 
limits $L_X\la 10^{40}$--$10^{41}$~erg~s$^{-1}$ \citep{Fornasini_2018}. The most 
hyper-luminous infrared galaxies with star-formation have
$L_X \sim 4 \times 10^{39}~ {\rm erg~s)}^{-1}\times ({\rm SFR}/ (\Msol~{\rm yr}^{-1})$
\citep{Mineo_2013}, although this relation could evolve with redshift 
\citep{wang2023}. These potential XRB contributions are negligible for the AGN identified in this work,  which have $L_X > 10^{42}$ erg s$^{-1}$, and therefore we 
ignore XRB emission and SFR contamination. 

44 NuSTAR sources lack an XMM counterpart (see Figure~\ref{fig:venn}), so \Zhao\ and \Silver\ do not report secure optical counterparts due to NuSTAR's coarser angular resolution compared to XMM; therefore, these sources are excluded from this multiwavelength analysis.


\subsection{Ultraviolet, Visible, and Infrared Spectrophotometry}

\begin{figure}
    \centering
    \hspace{-0.2cm}
    \includegraphics[width=0.99\linewidth]{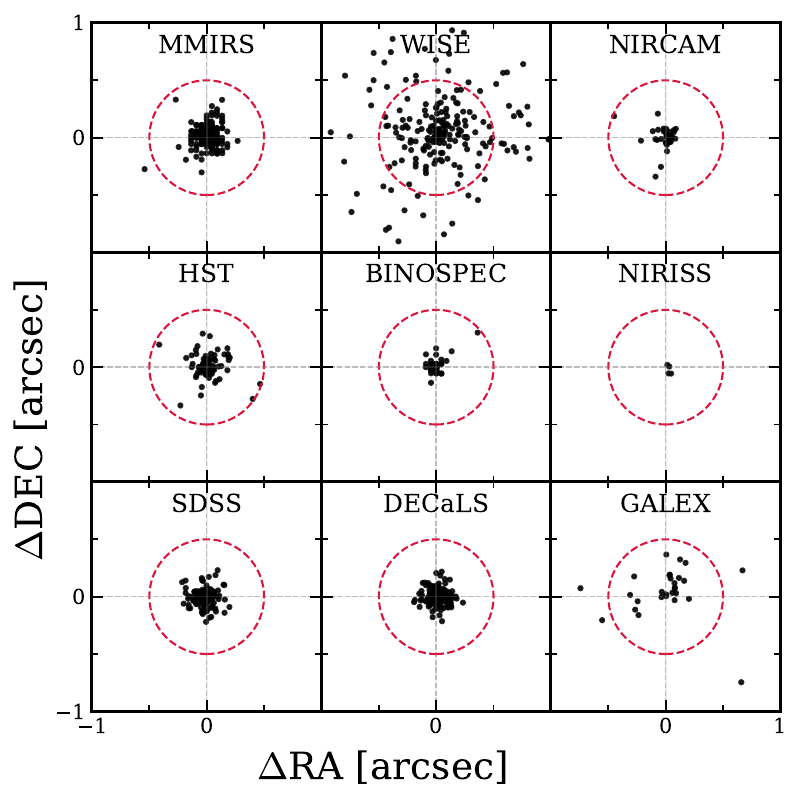}
    \caption{Position  offsets relative to HSC \textit{i} band imaging. The red-dashed circle has a radius of 0\farcs5. Each subplot lists the survey in text.}
    \label{fig:raDec}
\end{figure}

\begin{deluxetable}{lcccr}
\tablecaption{Catalog Matches from \Zhao\ and \Silver\ HSC Positions}
\tablewidth{0pt}
\tabletypesize{\scriptsize}
\tablehead{
  \colhead{SURVEY}          & \colhead{MATCHES}  & \colhead{Radius ['']} & \colhead{$N_{\rm FILTER}$} & \colhead{$\lambda$ [$\mu$m]} \\
  \colhead{(1)}             & \colhead{(2)}      & \colhead{(3)}         & \colhead{(4)}              & \colhead{(5)}
}
\startdata
GALEX            & 32                         & 3.0 & 2 & 0.15--0.23 \\
SDSS             & 121                        & 0.25 & 5 & 0.35--1.0 \\
HST/ACS/WFC3     & 114                        & 0.5 & 3 & 0.28--0.6 \\
DECaLS/$r$-band  & 220                        & 0.25 & 1 & 0.62 \\
MMT/Binospec     & 67$^{\tablenotemark{a}}$   & 0.5 & \nodata & 0.4--1.0 \\
MMT/MMIRS        & 157                        & 0.5 & 4 & 1.0--2.0 \\
JWST/NIRCam      & 37                         & 0.5 & 8 & 0.9--5.0 \\
JWST/NIRISS      & 4$^{\tablenotemark{b}}$    & 0.5 & 1 & 2.0 \\
WISE             & 228                        & 1.0 & 4 & 3.4--22 \\
\enddata
\tablenotetext{a}{The matching with Binospec returned 67 matches, of which only 21 were used downstream because they supplied new spectroscopic redshifts with quality flags $\geq 3$.}
\tablenotetext{b}{Four NIRISS sources were matched, though only two provided new spectroscopic redshifts.}
\label{tab:matches}
\tablecomments{Position match statistics to the HSC R.A. and Decl. reported in the \Zhao\ and \Silver\ catalogs of optical counterparts. Column (1) gives the survey name, column (2) gives the number of positional matches within the search radius, column (3). Columns (4) and (5) give the number of broadband filters and wavelength coverage of the survey.}
\end{deluxetable}

The multiwavelength coverage of the NEP TDF and counterpart matching were detailed by \Zhao\ and \Silver. This work ignores five identified stars and 105 X-ray sources with ``unsecure'' multiwavelength counterparts, leaving 343 extragalactic X-ray sources with secure counterparts. Table~\ref{tab:matches} lists the multiwavelength datasets available.

The primary ground-based dataset was from Subaru/HSC, which covered nearly the entire XMM field in
$g$, $i2$, and $z$ \citep{Oi2021, Taylor_2023} to  5$\sigma$  depth $\sim$26.7~mag in $i2$.
Additional wavelength and area coverage came from SDSS ($u$, $g$, $r$, $i$, $z$, $i\la24$) \citep{Abdurro'uf2022} and MMT/MMIRS ($Y$, $H$, $J$, $K$, $J\la23.5$) \citep{Willmer2023}. Additional $r$-band photometry ($r\la23.4$) came from the Legacy Survey Data Release 10 (LS DR10) queried from the tractor catalogs hosted by Astro Data Lab, which provides model-based photometry from the Beijing-Arizona Sky Survey (BASS)\null.
Only the \textit{r}-band observations were used because the  Subaru/HSC imaging covers the other bands at better depth and angular resolution. Matching was done within a 1\arcsec\ radius.

Archival GALEX observations are also available, with a single AIS and MIS observation covering part of the NEP with Target Names 	
AIS\_1\_1\_47 and MISDR1\_09937\_0350 with exposure times of 184 and 2705.25 seconds, respectively. The level 2 pipeline products from MAST identify which sources were within the field of view of these two observations. The MIS observation has a $5\sigma$  depth  $\sim$25.62 and $\sim$25.39 in NUV and FUV, respectively, with the AIS observation at $\sim$24.25 and $\sim$23.85 in NUV and FUV, respectively. Catalog photometry for GALEX was obtained from 2012GMSC..C...0000S and 2012GASC..C...0000S hosted at NED; matching was done within a 3\arcsec\ radius.

At longer wavelengths, space-based observations with WISE (3.4, 4.6, 12, 22~\micron) \citep{Schlafly2019} cover the whole field. Sensitivity was $\sim$21.5~mag in W1 and less in the longer wavelengths.

HST (F275W, F435W, F606W) (\citet{Obrien2024}, Jansen et al.\ in prep.)\ observations with depths to 29.5~mag (2$\sigma$) in F606W are available for the central area of the XMM field.  JWST (F090W, F115W, F150W, F200W, F277W, F356W, F410M, F444W) (\citet{Windhorst2023}, Jansen et al. in prep.)\  coverage is limited to four ``spokes'' within the HST area and reaches 29~mag ($5\sigma$) in F200W\null. 
The HST and JWST data were drizzled to 30 milli-arcseconds pixel scale and matched to GAIA DR3 \citep{Gaia2023}. Catalogs are not yet published; therefore, we produced catalogs with \texttt{SourceExtractor} \citep{Bertin1996} on the mosaics described  \citep{Obrien2024, Windhorst2023}.  Measurements were in dual-image mode  using F444W as the detection band whenever possible and F606W otherwise. Detection required 9 contiguous pixels atleast 1.5$\sigma$ above the background, 32 deblending sub-thresholds with a 0.001 minimum contrast for object deblending. Automated position matches within 0\arcspt5 were confirmed by visual inspection of the \texttt{SourceExtractor} segmentation map against Subaru/HSC positions . This process identified XMM-ID 268, identified by \Zhao\ as a bright star, with a red and compact AGN candidate missed by the HSC data.  XMM-ID 230 was HSC-identified as a low-surface-brightness spiral and is near two bright and red AGN candidates missed by HSC\null, though we treat the nearest object to the HSC-identified spiral as the AGN candidate. SED fits  using only NIRCam data show that both XMM-ID 230 and 268 are likely AGN, and we treat those sources as the X-ray emitter.

Figure~\ref{fig:NEP} illustrates the difference in sky coverage for these observations; notably, most X-ray identified counterparts are with Subaru/HSC and/or MMT/MMIRS. For sources with overlapping detections in the $u$/$g$/$r$/$i$/$z$ bands, we favor HSC detections. SDSS $ugriz$ observed magnitudes are corrected for galactic extinction following \citet{Schlafly2011} at the NEP TDF coordinates, which already have low Galactic extinction to begin with \citep{Jansen_2018}. Table~\ref{tab:matches} tabulates the position matching of HSC positions with the extant catalogs, along with their statistics and separations.

\Zhao\ and \Silver\ provide spectroscopic redshifts for 80 sources.  In addition, 21 redshifts are available from new MMT/Binospec observations (Willmer et al.\ in prep.) and two from JWST/NIRISS  (Estrada-Carpenter et al.\ in prep.).
The DESI DR1 catalog \citep{DESICollaboration2026} gave two additional spectroscopic redshifts, and SDSS in the Dark Energy Camera Legacy Surveys (DECaLS) \citep{Dey2019} gave one more. Positional matches were within 0\arcspt5 except for DESI DR1, which used 0\farcs7. New spectroscopic redshifts are identified via footnote and flag in Table~\ref{table:z_fAGN_class} of the Appendix.

\subsection{Radio and Sub-mm}
High‐resolution 3~GHz imaging of the NEP TDF was obtained with the Karl G.\ Jansky Very Large Array with full details on the data presented by \citet{Hyun2023}.
Radio counterparts of X-ray sources in the field have already been identified with the help of HSC \citep{Hyun2023} (or JWST positions for sources covered by NIRCam; \citep{Willner2026}). 
We did not include the radio flux density in the SED fitting because one detection does not well constrain the synchrotron emission, and because tests showed that including this flux did not improve the SED fit quality.
 
The NEP TDF was surveyed at 850$\mu$m with the SCUBA-2 camera on the James Clerk Maxwell Telescope for 41.3 total hours. Typical positional uncertainty is 13\arcsec, but some sources have VLA counterparts that decrease the uncertainties to 0\farcs9 \citep{Hyun2023}. For X-ray sources consistent with a SCUBA-2 position, we used the 850$\mu$m flux density as an upper limit in the SED fitting.  For other sources covered by SCUBA-2, the 850~\micron\ upper limit was set to 5~mJy.  Counterparts to X-ray sources were identified within \Zhao\ and \Silver\ catalogs via VLA source ID.

\section{SED Fitting \& Redshifts}
\label{sec:methods}
\subsection{Techniques}
\label{sec:techniques}

We use the \texttt{CIGALE} SED fitting tool \citep{Boquien2019, Yang2020, Yang2022} due to its robust galaxy SED modeling, included AGN templates, and the ability to fold in X-ray fluxes into the SED fit. \texttt{CIGALE} is a public code that leverages energy balance from the ultraviolet to the far-infrared to fit the SEDs of galaxies and infer physical properties of the host-galaxy and of individual components. The motivation behind the SED analysis is to incorporate AGN templates and galaxy components (e.g., simple stellar populations (SSPs), dust attenuation and emission, nebular emission) to infer the host-galaxy properties of these X-ray sources.

We design \texttt{CIGALE} with parameters that define the SED template construction, tabulated in Table~\ref{tab:sedparams-custom} of the Appendix. Parameters listed in this table are only those that differ from the default values. \texttt{CIGALE} builds SEDs through the combination of user-defined components such as SSPs, an initial mass function (IMF), dust attenuation prescriptions, nebular gas emission, and AGN emission. We use the most recent version of \texttt{CIGALE} (2025.1), which is built using additional SSP templates and libraries from Charlot \& Bruzal (CB19), a modern extension of the popular\citet{Bruzual_2003} templates\setcounter{footnote}{0}
\footnote{The \texttt{CIGALE} GitHub explains that CB19 templates can be incorporated and built into the code, though the data files are too large to be included in the \texttt{git} and must be added separately, {\href{https://gitlab.lam.fr/cigale/cigale/-/commit/79488c560dbef388600c587f929ec439980108f6}{https://gitlab.lam.fr/cigale/cigale/-/commit/79488c560dbef388600c587f929ec439980108f6}}.}. We modeled the star-formation history (SFH) with the stochastic SFH module from \citet{Carvajal2025}, in which a delayed-$
\tau$ parametric SFH is multiplicatively modulated by a stochastic process with a temporal structure defined by a broken power law. The stochastic component is controlled by the normalization of the power spectrum density ($\sigma$) (i.e., the ``burstiness" level), the correlation timescale for starbursts ($\tau_{\rm break}$), and the high frequency power-law slope of the power spectrum density, which governs power at longer timescales ($\alpha$). We generate $N_{\rm SFH}$ stochastic realizations with random seeds to better sample different burst patterns while balancing computational cost. These SFHs, combined with the CB19 SSP's and the Chabrier IMF, undergo dust attenuation and reprocessed emission to produce the stellar and dust components to the SED \citep{Calzetti1994, Dale2014}. The \texttt{SKIRTOR} AGN model \citep{Stalevski2016} is used to produce the AGN component to the SED. The parameters of interest include the opening angle of the dusty torus, the inclination (i.e., viewing angle) of the disk, and the total contribution of the AGN emission to the total SED from $0.1-30\mu m$ ($f_{\rm AGN}$). \citet{Lopez2024} X-ray models are used to produce AGN emission from the corona and hot accretion disk. \texttt{CIGALE} requires that the input X-ray fluxes are absorption corrected, which follows from the methods of \citep{Creech2025} and uses \texttt{xspec} to generate intrinsic X-ray fluxes for the sources.  A redshift grid of 500 points uniformly spaced in $\log(1 + z)$ over $0.1 \leq z \leq 5$ is adopted, corresponding to a constant resolution of $\Delta z / (1 + z) \approx 0.004$.

Meaningful SED fits require minimum of four visible--infrared data points.
For the 261 sources meeting this criterion, we ran \texttt{CIGALE} twice; one run with only ultraviolet, visible, and infrared photometry in order to best anchor photometric redshifts for sources lacking spectroscopic redshifts, and a second run with X-ray, ultraviolet, visible, infrared, and sub-mm upper limits fixed at the source's best redshift solution from the first run. This approach leverages the useful photometry for photometric redshifts while reducing computational cost when fitting SEDs across the entire electromagnetic spectrum. The second run's physical inferences are used for all analysis of galaxy properties.

\subsection{Quality Control}

\label{sec:samplequality}

\begin{figure} 
    \centering
    \hspace{-0.2cm}
    \includegraphics[width=0.95\linewidth]{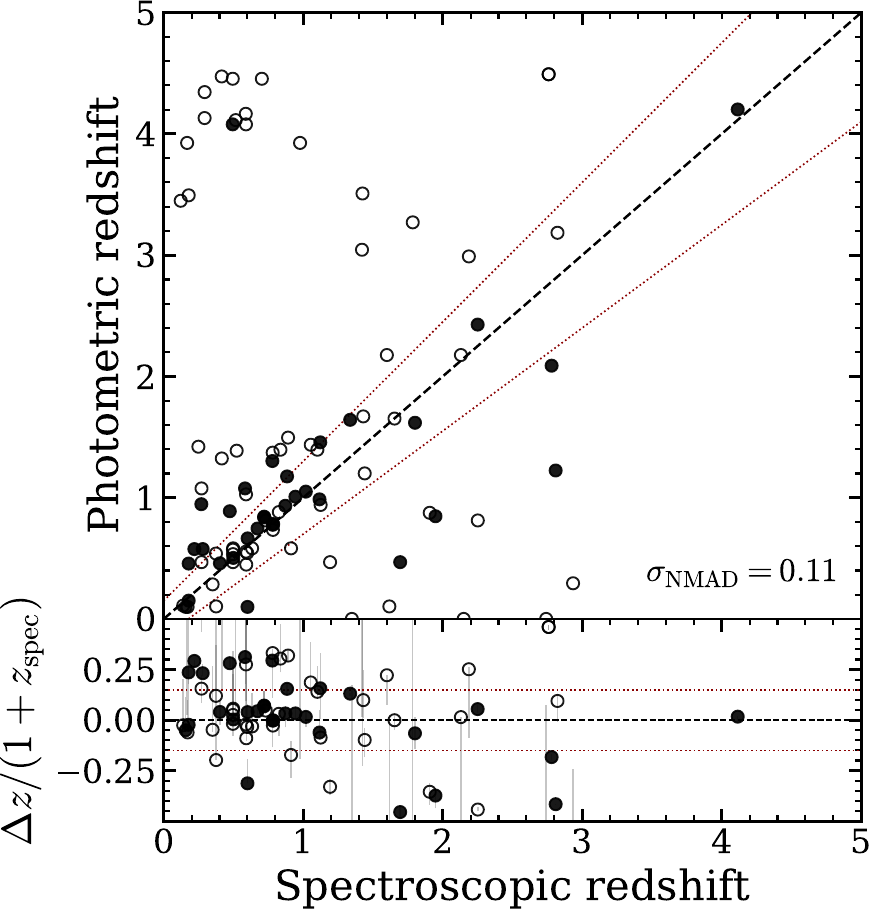}


    \caption{Best-fit photometric redshift via \texttt{CIGALE} versus the known spectroscopic redshift. Filled points require $Q_z < 2$ and 11 broadband detections; the redshift uncertainty for those 33 sources is characterized by $\sigma_{\rm NMAD}=0.11$. Residual offsets normalized to $\log(1+z)$ are shown below.}
    \label{fig:SpeczPhotz}
\end{figure}


For the first \texttt{CIGALE} run with ultraviolet, visible, and infrared photometry,  we aim to gauge our photometric redshift reliability. To test this, we compute the best-fit photometric redshift with \texttt{CIGALE} and compare to the known spectroscopic redshift when available. This reliability is captured in Figure~\ref{fig:SpeczPhotz}. About two-thirds of the sample has optical and infrared coverage with six or more bands, and many objects have deep coverage with JWST and/or HST. The redshift precision for filled points---which are objects with ample broadband detections and well defined photometric redshifts---is $\sigma_{\rm NMAD} = 0.11$, consistent with other analyses having this degree of wavelength coverage for AGN \citep{Burlon2011}. However, some objects have poor redshift reliability due to having only four broadband filters or poor wavelength coverage of key rest-frame features in their SEDs (e.g., Balmer break, Lyman break). These SED features lead to some degenerate redshift solutions   apparent in the $P(z)$ curves in Figure~\ref{fig:zhist}. In Figure~\ref{fig:SpeczPhotz}, there is one filled point with a low spectroscopic redshift that was best fit at $z>4$, and visual inspection of this SED reveals a dramatic Balmer break that causes this failed solution.

To further assess the reliability of photometric redshifts, we adopt the $Q_z$ parameter from \citet{Brammer2008}, which gauges how well the redshift probability curve, $P(z)$, is constrained with respect to the redshift grid. The dimensionless redshift quality $Q_z$ is defined as:
\begin{equation}
Q_z \equiv \frac{\chi^2}{N_{\mathrm{filt}} - 3}
\frac{z_{\mathrm{up}}^{99} - z_{\mathrm{lo}}^{99}}{p_{\Delta z = 0.2}} \quad,
\end{equation}
where $\chi^2$ is for the best template fit, \( N_{\mathrm{filt}} \) is the number of filters used in the fit, \( z^{99}_{\mathrm{up}} \) and \( z^{99}_{\mathrm{lo}} \) are the 99\% upper and lower redshift confidence limits, and \( p_{\Delta z = 0.2} \) is the posterior probability density integrated within \( \Delta z = 0.2(1+z) \) around the photometric redshift.  Redshifts with $Q_z\lesssim2$ are reliable and have well-peaked $P(z)$ distributions, while larger \( Q_z \) indicates a less reliable photometric redshift estimate. Values $\lesssim10$ are broad and/or bimodal, while values $>10$ are unreliable and should be flagged. These $Q_z$ values categorize the SED fits as high-fidelity (Category A: $Q_z<2$, \( N_{\mathrm{filt}}\ge8 \) including near-infrared coverage), fair (Category B: $Q_z<10$, \( N_{\mathrm{filt}}\ge6 \) including near-infrared coverage), and bad fits (Category C: all others). 
Objects with a spectroscopic redshift were classified as Category A\null. For the sample of 261 objects, 127 are Category A (102 with a spectroscopic redshift), 115 are Category B, and 19 are flagged as Category C.
Figure~\ref{fig:SpeczPhotz} shows the distributions of these categories with redshift, and
example SEDs are shown in Figure~\ref{fig:SED}.

\subsection{Redshift Distribution}

The final photometric redshifts were chosen depending on fit category.  For Category~A, these were the $P(z)$-weighted average redshift solution (\texttt{bayes.universe.redshift} from the \texttt{CIGALE} output).  For Category~B, the maximum-likelihood (i.e., peak $P(z)$) redshift was used, and for Category~C, \texttt{CIGALE}s \texttt{best.universe.redshift} solution, i.e., the redshift of the SED fit with the lowest $\chi^2$ was used.  These redshifts were then kept fixed in the second \texttt{CIGALE} run, where the full ensemble of X-ray, ultraviolet, visible, infrared, and sub-mm data were fit. 

The redshift distribution of sources, shown in Figure~\ref{fig:zhist},  peaks at $z\sim 1$, consistent with other X-ray surveys \citep{Aird2015, Civano_2016, Marchesi2016}. 
The $P(z)$ curves identify where the most likely redshift solutions exist across the redshift grid. The general shape of the redshift distribution is shown by the cumulative $P(z)$ curve smoothed by a Gaussian kernel with $\sigma=\Delta\,z/(1+z)=0.15$ across a $\log\,(1+z)$ redshift grid.
For Category~A and B sources, the median redshift is $\langle z\rangle=0.98$, the average redshift is $\bar z=1.34$, and the maximum redshift is $4.68$. 

\begin{figure}
    \centering
    \hspace{-0.2cm}
    \includegraphics[width=0.99\linewidth]{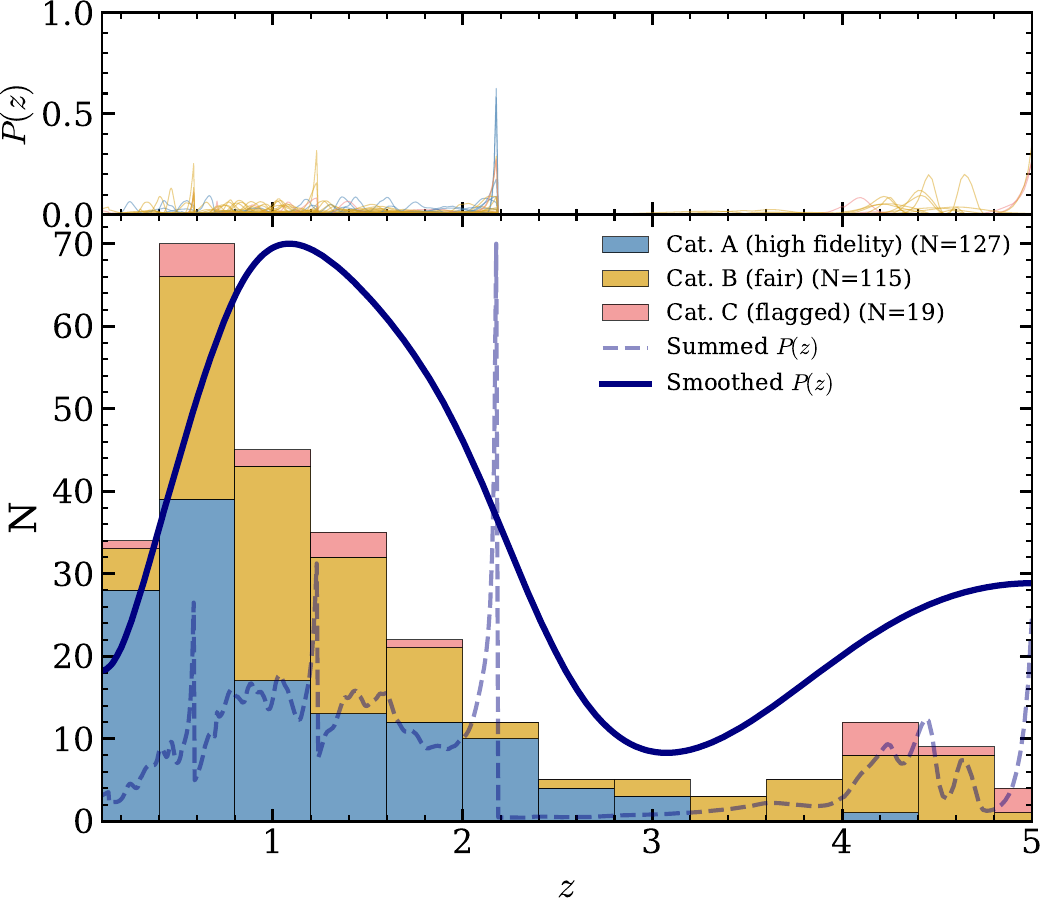}
    \caption{Redshift distribution for the sample.  Categories A, B, and C are shown as blue, yellow, and red histograms, respectively. The top panel shows the individual $P(z)$ curves for all sources. In the main panel, the navy blue dashed line shows the sum of all $P(z)$, and the solid navy blue line shows the summed distribution smoothed by $
    \log(1+z)=0.15$.}
    \label{fig:zhist}
\end{figure}

\section{Results}
\label{sec:results}

\begin{figure*}
    \centering
    \hspace {-0.2cm}
    \includegraphics[width=0.49\textwidth,clip=true,trim=0 31 0 0]{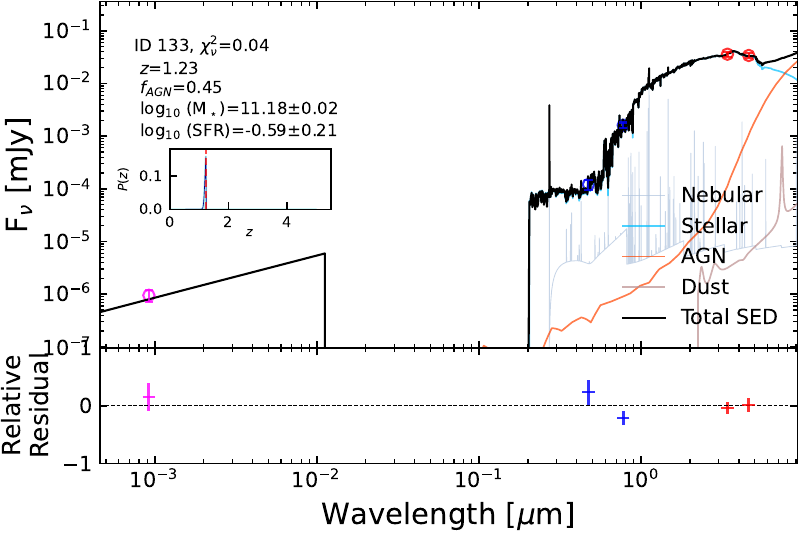}
    \includegraphics[width=0.49\textwidth,clip=true,trim=0 31 0 0]{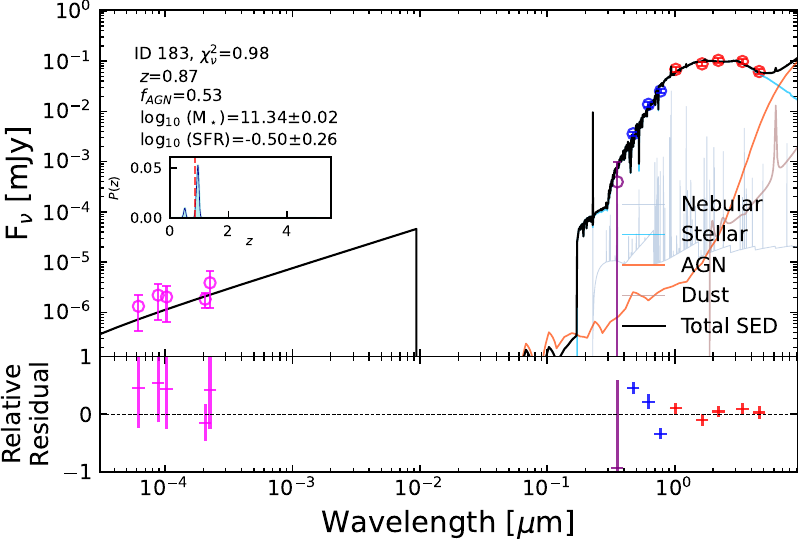}
    
    \centering
    \hspace {-0.2cm}

    \includegraphics[width=0.49\textwidth,clip=true,trim=0 31 0 0]{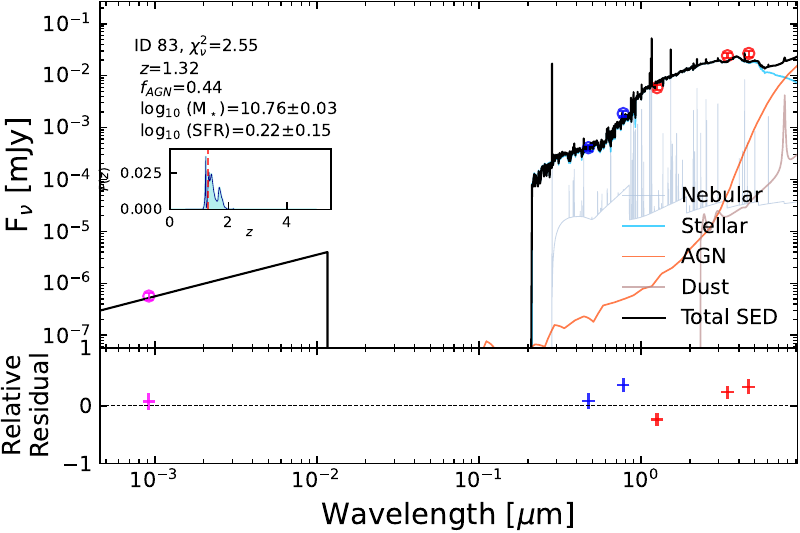}
    \includegraphics[width=0.49\textwidth,clip=true,trim=0 31 0 0]{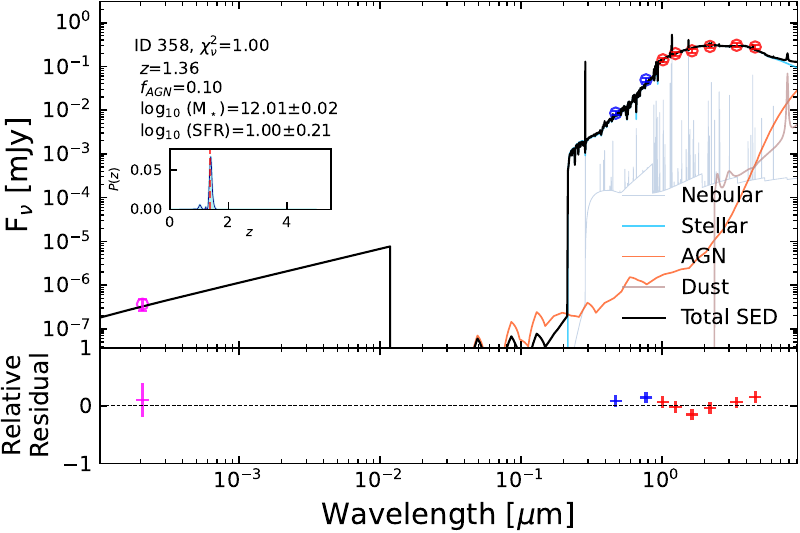}
    
    \centering
    \hspace {-0.2cm}

    \includegraphics[width=0.49\textwidth]{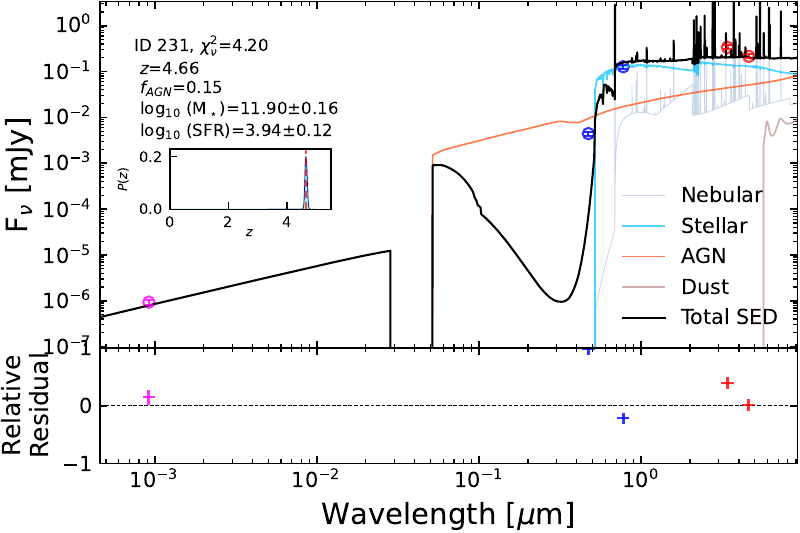}
    \includegraphics[width=0.49\textwidth]{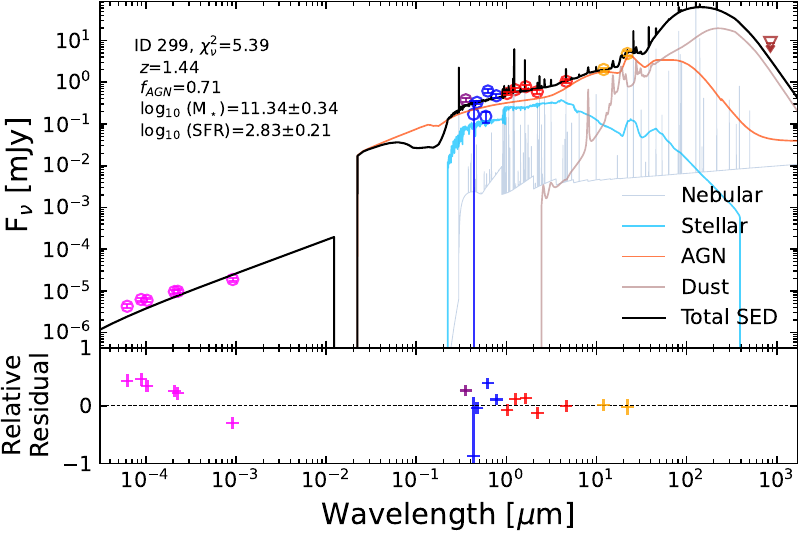}

    \caption{Example SED fits from \texttt{CIGALE}\null. The XMM ID, goodness of fit metric, $f_{\rm AGN}$, $M_*$, and SFR are listed in the top left of each panel. The $P(z)$ curve is shown as an inset figure if the source lacks a spectroscopic redshift. Colored lines show the individual SED components as indicated in each panel, and the black line shows their sum as the total SED. Points show the observed data colored by observed wavelength. Fit residuals are shown in the lower part of each panel. These six example SEDs are representative of the full sample demographics in regard to filter coverage and galaxy type.}
    \label{fig:SED}
\end{figure*}

The SED fitting across the X-ray, ultraviolet, visible, infrared, and sub-mm wavelengths enables robust inferences on both the AGN and its host galaxy. Galaxy properties include stellar mass ($M_*$), star-formation rate (SFR), and offset from the star-forming main sequence SFR ($\Delta\rm SFMS$).  Properties related to the black hole are AGN luminosity ($L_{\rm AGN}$), offset from the population-average black-hole accretion rate ($\Delta\rm BHAR$), and fraction of the 0.1--30~\micron\ luminosity from the AGN ($f_{\rm AGN}$)\null. These put the X-ray hosts into the context of the wider galaxy population to identify relationships between SMBH activity and galaxy evolution.

\subsection{The Star-Forming Main Sequence}
\label{s:sfms}
Immediate properties of interest about the host galaxy from the SED fitting are $M_*$ and SFR\null.
These are related by the SFMS, defined by \citet{Popesso2023} as
\begin{eqnarray}
\log \mathrm{SFR}_{\max}(t) &=& 2.693 - 0.186\,t \\
\log M_0(t) &=& 10.85 - 0.0729\,t \\
\log \mathrm{SFR_{\rm MS}}(M_\star, t) &=& \log \mathrm{SFR}_{\max}(t) \nonumber\\
&& - \log\!\bigl[1 + (M_\star/M_0(t))^{-0.99}\bigr] ~,
\end{eqnarray}
where $t(z)$ is the age of the universe in Gyr, $\rm SFR_{max}$ is the normalization of the peak SFR on the main sequence, and $M_0$ is the characteristic stellar mass at which the main sequence begins to turn over. 
The offset from the SFMS is then
\begin{equation}
    \Delta{\rm SFMS} = \log \mathrm{SFR} - \log \mathrm{SFR_{\rm MS}}(M_*,t(z))\quad,
\label{eq:dsfms}
\end{equation}
where $t$ is calculated from $z$ and the adopted cosmology.
These offsets are shown in Figure~\ref{fig:sfms}. 

Most of the X-ray host galaxies reside below the SFMS, with some fully quenched as defined by $\Delta{\rm SFMS}<-1.5$ \citep{Renzini2015}. At the highest $L_{\rm AGN}$, the galaxy sample begins to preferentially reside at or above the SFMS, with some suggesting starburst activity. Lower $L_{\rm AGN}$ host-galaxy SEDs are less AGN dominated, while lower $M_*$ host galaxy SEDs are more AGN dominated. The stellar masses of these host galaxies---especially at the high-mass regime (i.e., $M_*>10^{10.5} M_\odot$)---span the full range of $\Delta{\rm SFMS}$ and $f_{\rm AGN}$. The low-mass galaxies ($\log(M_*/M_\odot)\lesssim 10$) preferentially exist near the SFMS. 

The specific AGN luminosity ($L_{\rm AGN}/M_*$), which can serve as a proxy for Eddington ratio \citep[see][or the crude conversion in Figure~\ref{fig:agnHist}]{Aird2018, Alexander2025},  reveals a strong positive correlation with $\rm\Delta SFMS$. The partial Spearman rank correlation coefficient ($\rho_{\rm partial}$) is computed as $+0.73$ between the specific AGN luminosity and the SFMS offset, controlling for $M_*$, at probability $p<0.001$. As specific AGN luminosity increases, galaxies reside closer to the SFMS. In other words, the star-forming host galaxies in the sample are coincident with the most efficiently accreting SMBHs. The low-mass galaxies are among the most specifically luminous AGN and preferentially exist near or above the SFMS, whereas the more massive galaxies represent SMBHs within galaxies below the SFMS that accrete far below the Eddington limit.


\begin{figure*}
    \hspace*{-0.8cm}
    \centering
    \includegraphics[width=0.75\linewidth]{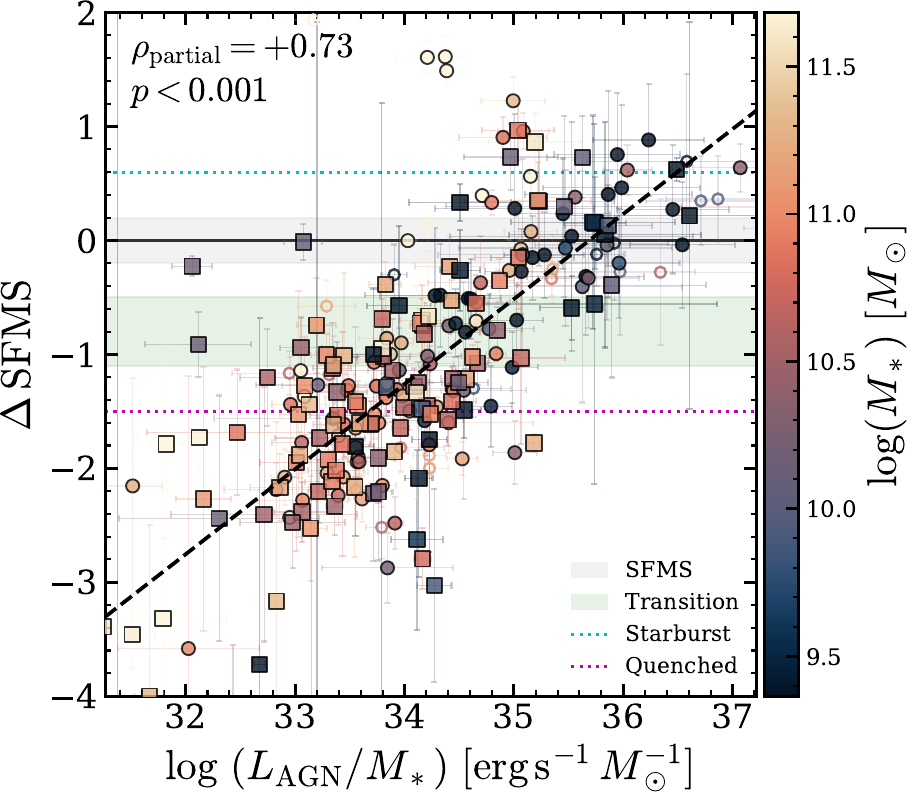}
    \includegraphics[width=0.99\linewidth]{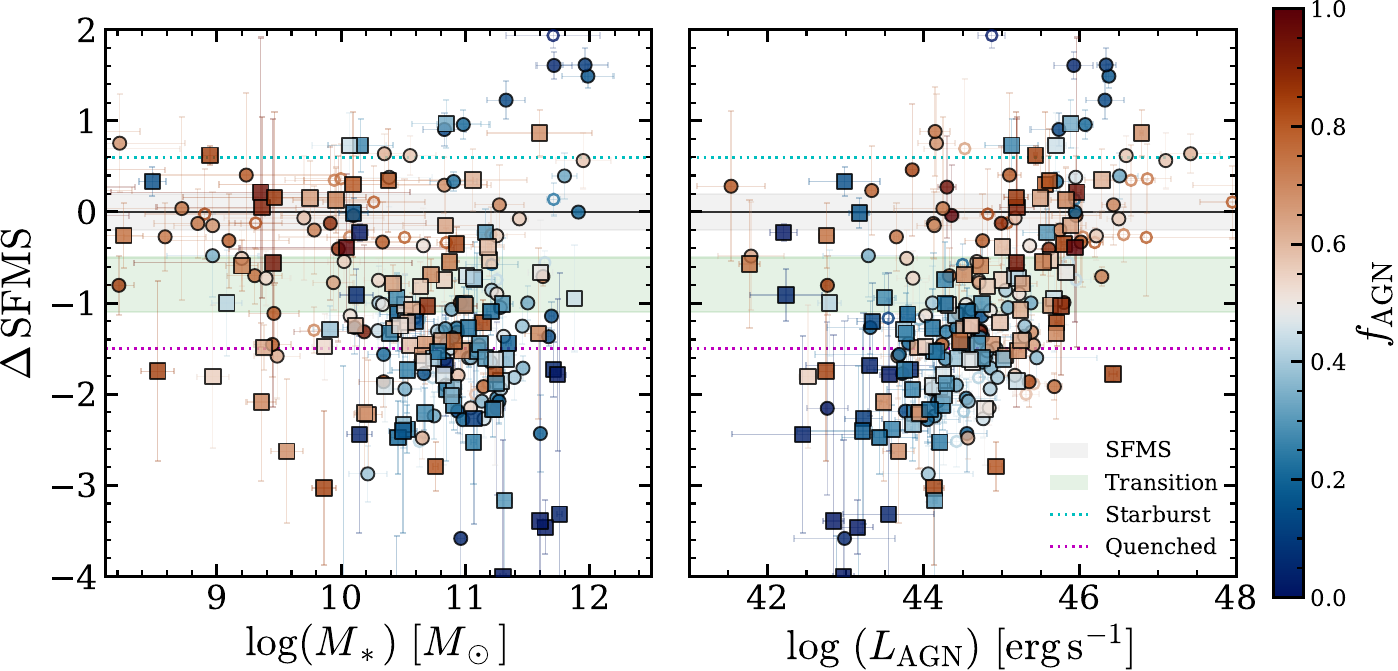}

    \caption{$\rm\Delta{SFMS}$ (Equation~\ref{eq:dsfms}) vs.\ galaxy properties.  Filled squares, filled circles, and open circles identify galaxies with category A, B, and C SED fits, respectively. Horizontal lines identify the starburst (teal) \citep{Rodighiero2011}, SFMS (grey), transition (green), and quenched boundaries (magenta) \citep{Renzini2015}. \textit{Top:}  $\rm\Delta{SFMS}$ vs.\ specific AGN luminosity, i.e.,  $L_{\rm AGN}/M_*$. Point colors indicate $M_*$ according to the color bar at right. The partial Spearman rank correlation coefficient ($\rho_{\rm partial}$) between $L_{\rm AGN}/M_*$ and $\rm \Delta SFMS$, controlling for $M_*$ is shown at top left: $p<0.001$ indicates the probability of obtaining the correlation under the null hypothesis of no correlation.  \textit{Bottom left:} $\rm\Delta{SFMS}$ vs.\  $M_*$. \textit{Bottom right:} $\rm\Delta{SFMS}$ vs.\ $L_{\rm AGN}$. $L_{\rm AGN}$ was computed using the \citet{Duras2020} $k_X$ luminosity-dependent bolometric correction factor to $L^{\rm 2-10~keV}_X$ from \texttt{CIGALE} (Equation~\ref{eq:kbol}).
 In both lower panels, point color indicates $f_{\rm AGN}$ according to the color bar at right.}
    \label{fig:sfms}
\end{figure*}

\subsection{Morphologies}

The sample subset that has high-resolution imaging from HST and JWST can be additionally examined through their morphologies. These are only a minority of the X-ray sample, making a full quantitative analysis of galaxy morphologies impossible. However, the broadband imaging from 0.2--4.5~\micron\ gives a qualitative sense.

Figure~\ref{fig:RGB} shows all but two of the sources with overlapping HST and JWST coverage. The morphologies of these galaxies are diverse. Some galaxies are small and faint (ID 225/245/256). Their X-ray luminosities are at the lower end of the sample distribution, suggesting less massive yet growing SMBHs. Some galaxies display distinct spiral arm structure (ID 252/222/122), while other disc-like galaxies suggest tidal-force distortions to their morphologies (ID 224/269/268). A handful of galaxies sport red diffraction spikes from their centers (ID 213/163/222/327) reported by \citet{Ortiz2024} as resolved AGN emission. There are even signs of galaxy groups and mergers (ID 230/265/285). A star is also shown (ID 237). In general, the galaxy morphologies of these X-ray AGN are disc-like rather than those of early-type elliptical galaxies.

\begin{figure*}
    \centering
    \includegraphics[width=0.17\linewidth]{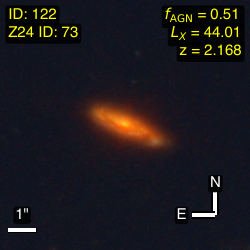}%
    \includegraphics[width=0.17\linewidth]{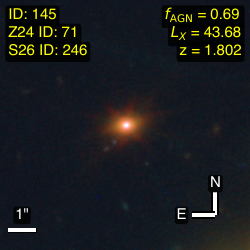}%
    \includegraphics[width=0.17\linewidth]{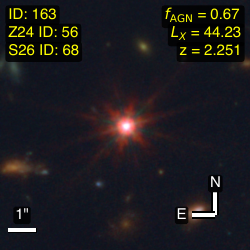}%
    \includegraphics[width=0.17\linewidth]{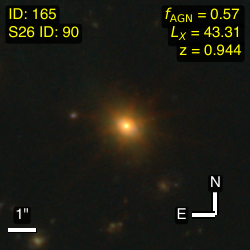}%
    \includegraphics[width=0.17\linewidth]{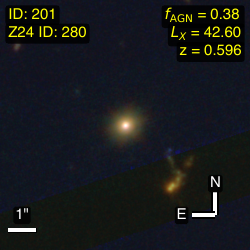}\\
    \includegraphics[width=0.17\linewidth]{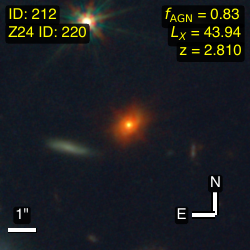}%
    \includegraphics[width=0.17\linewidth]{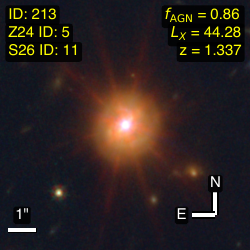}%
    \includegraphics[width=0.17\linewidth]{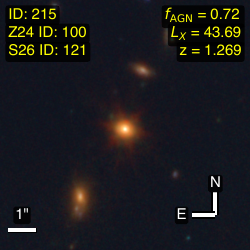}%
    \includegraphics[width=0.17\linewidth]{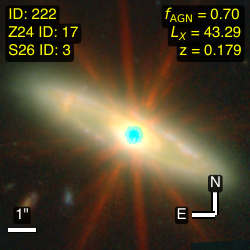}%
    \includegraphics[width=0.17\linewidth]{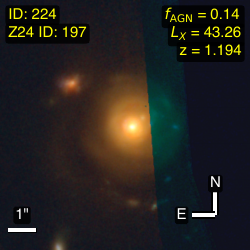}\\
    \includegraphics[width=0.17\linewidth]{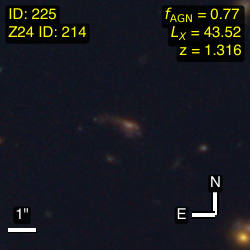}%
    \includegraphics[width=0.17\linewidth]{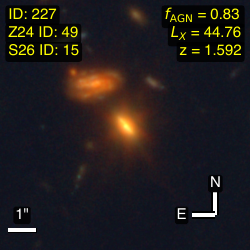}%
    \includegraphics[width=0.17\linewidth]{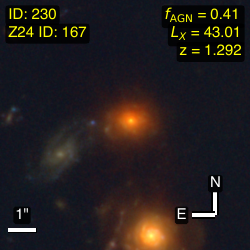}%
    \includegraphics[width=0.17\linewidth]{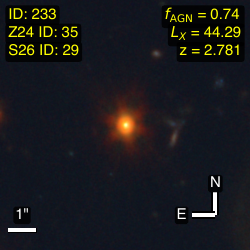}%
    \includegraphics[width=0.17\linewidth]{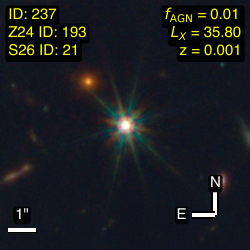}\\
    \includegraphics[width=0.17\linewidth]{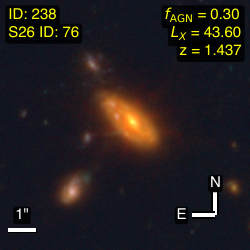}%
    \includegraphics[width=0.17\linewidth]{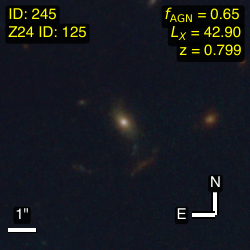}%
    \includegraphics[width=0.17\linewidth]{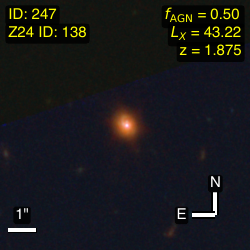}%
    \includegraphics[width=0.17\linewidth]{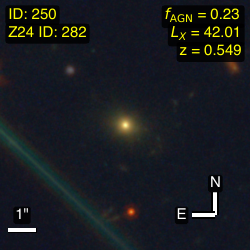}%
    \includegraphics[width=0.17\linewidth]{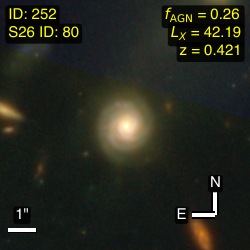}\\
    \includegraphics[width=0.17\linewidth]{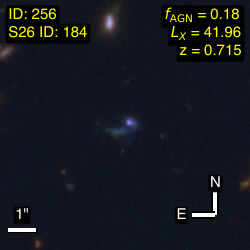}%
    \includegraphics[width=0.17\linewidth]{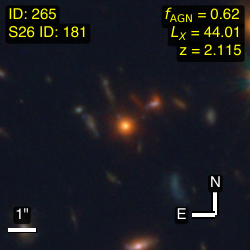}%
    \includegraphics[width=0.17\linewidth]{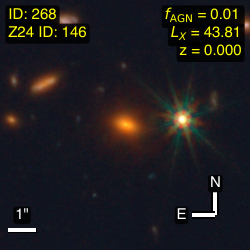}%
    \includegraphics[width=0.17\linewidth]{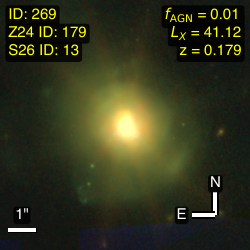}%
    \includegraphics[width=0.17\linewidth]{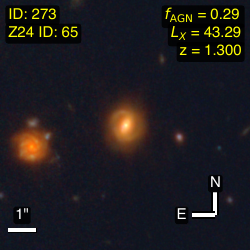}\\
    \includegraphics[width=0.17\linewidth]{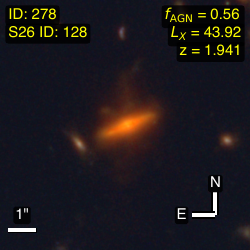}%
    \includegraphics[width=0.17\linewidth]{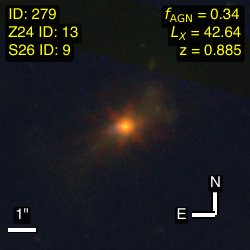}%
    \includegraphics[width=0.17\linewidth]{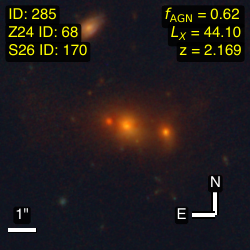}%
    \includegraphics[width=0.17\linewidth]{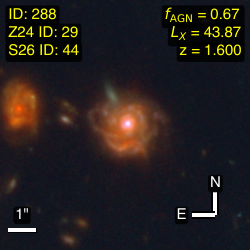}%
    \includegraphics[width=0.17\linewidth]{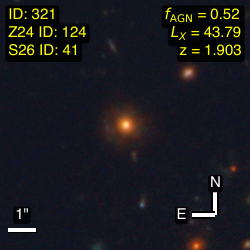}\\
    \includegraphics[width=0.17\linewidth]{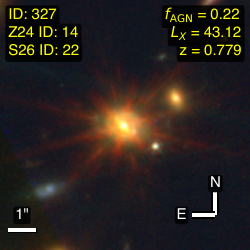}%
    \includegraphics[width=0.17\linewidth]{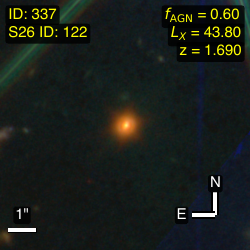}%
    \includegraphics[width=0.17\linewidth]{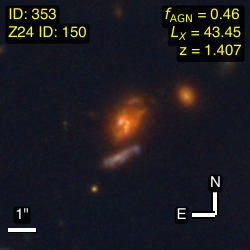}%
    \includegraphics[width=0.17\linewidth]{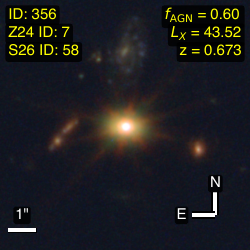}%
    \includegraphics[width=0.17\linewidth]{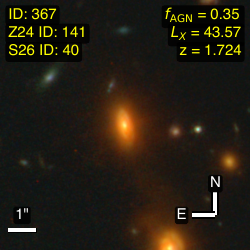}
    \caption{False color images using JWST+HST filters. Image colors are rendered as done in \citet{bowling2026}: R = F444W+F410M+F356W, G = F277W+F200W, B = F150W+F115W+F090W+F606W+F435W+F275W. IDs from this work, \Zhao, and \Silver\ are in the top left of each stamp, and physical properties $f_{\rm AGN}$, $L_X^{2-10\rm~keV}$, and $z$ are in the top right. The compass and angular scale are at the bottom corners.}
    \label{fig:RGB}
\end{figure*}

\subsection{The Central Engine}

\begin{figure}
    \centering
    \hspace{-0.2cm}
\includegraphics[width=0.99\linewidth]{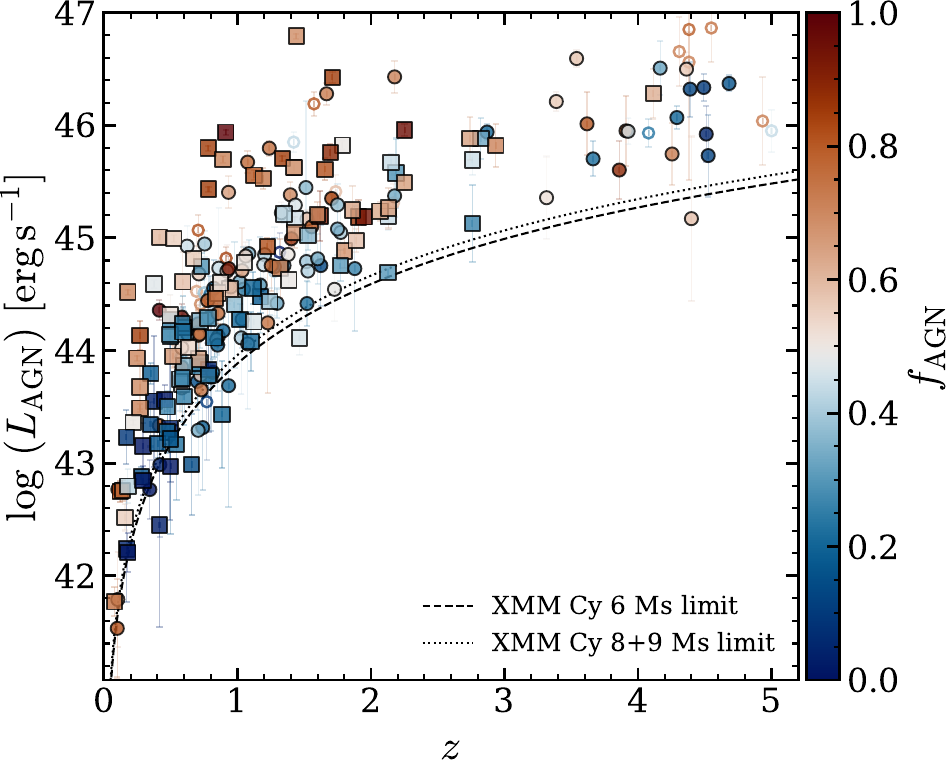}
    \caption{AGN luminosity versus redshift. Filled squares, filled circles, and open circles correspond to categories A, B, and C fits, respectively. Each point's color indicates  $f_{\rm AGN}$ according to the color scale at right. The black dashed lines correspond to the XMM survey luminosity limit for each cycle's observations.}
    \label{fig:LxVSz}
\end{figure}


The AGN luminosity can be derived from the X-ray emission as
\begin{equation}
\begin{aligned}
    L_{\rm AGN} &= K_{X}(L_{X}^{2-10\rm keV})  \\
    &= a \left[
        1 +
        \left(
        \frac{\log\left(L_{X}^{2-10\rm keV}/L_\odot\right)}{b}
        \right)^{c}
        \right],
\end{aligned}
\label{eq:kbol}
\end{equation}
where $K_X$ is the luminosity-dependent bolometric correction factor applied to $L_X^{\rm 2-10keV}$ from the SED fitting, and \textit{a}, \textit{b}, and \textit{c} are 15.33, 11.48, and 16.20, respectively \citep[][their Eq.~3]{Duras2020}. 
Figure~\ref{fig:LxVSz} shows the $L_{\rm AGN}$ distribution for the sample.
Equation~\ref{eq:kbol} also gives 
the luminosity limits of the XMM observations, shown in Figure~\ref{fig:LxVSz}:
\begin{align}
L_{\rm X,lim}^{\rm Cy\,6}(z)
&=
4\pi d_L^2(z)
\left(1.60\times10^{-14}\right)
(1+z)^{\Gamma-2}~,
\\[4pt]
L_{\rm X,lim}^{\rm Cy\,8+9}(z)
&=
4\pi d_L^2(z)
\left(1.90\times10^{-14}\right)
(1+z)^{\Gamma-2}~,
\end{align}
where $d^2_L$ is the luminosity distance, and $\Gamma=1.8$.

At $z\simeq1$--2, the sample galaxies are preferentially AGN-dominated. Low-luminosity AGN are sampled only at $z<1$, and some   $z>4$ sources exist, though not all of the photometric redshifts are reliable. The median $f_{\rm AGN}$ values by redshift bin are:
$0 \leq z < 1$: $\langle f_{\rm AGN} \rangle = 0.34 \pm 0.03$;
$1 \leq z < 2$: $\langle f_{\rm AGN} \rangle = 0.56 \pm 0.02$;
$2 \leq z < 3$: $\langle f_{\rm AGN} \rangle = 0.59 \pm 0.03$;
$z \geq 3$: $\langle f_{\rm AGN} \rangle = 0.39 \pm 0.07$.

\subsubsection{Black Hole Accretion Rate}

\begin{figure*}
    \centering
    \includegraphics[width=0.65\linewidth]{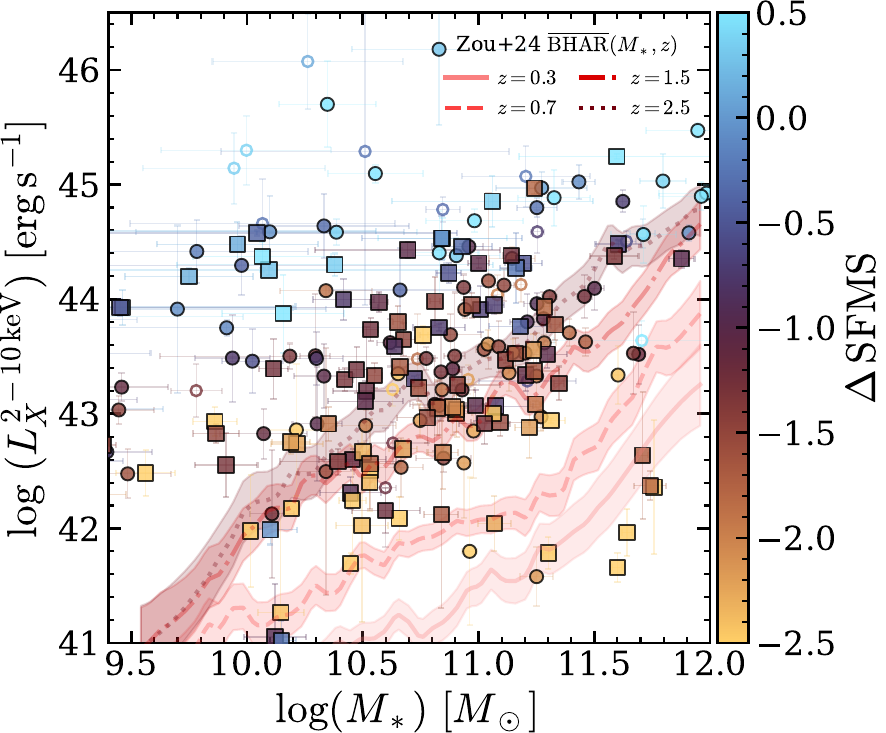}\\
    \hspace{-0.2cm}
    \centering
    \includegraphics[width=0.99\linewidth]{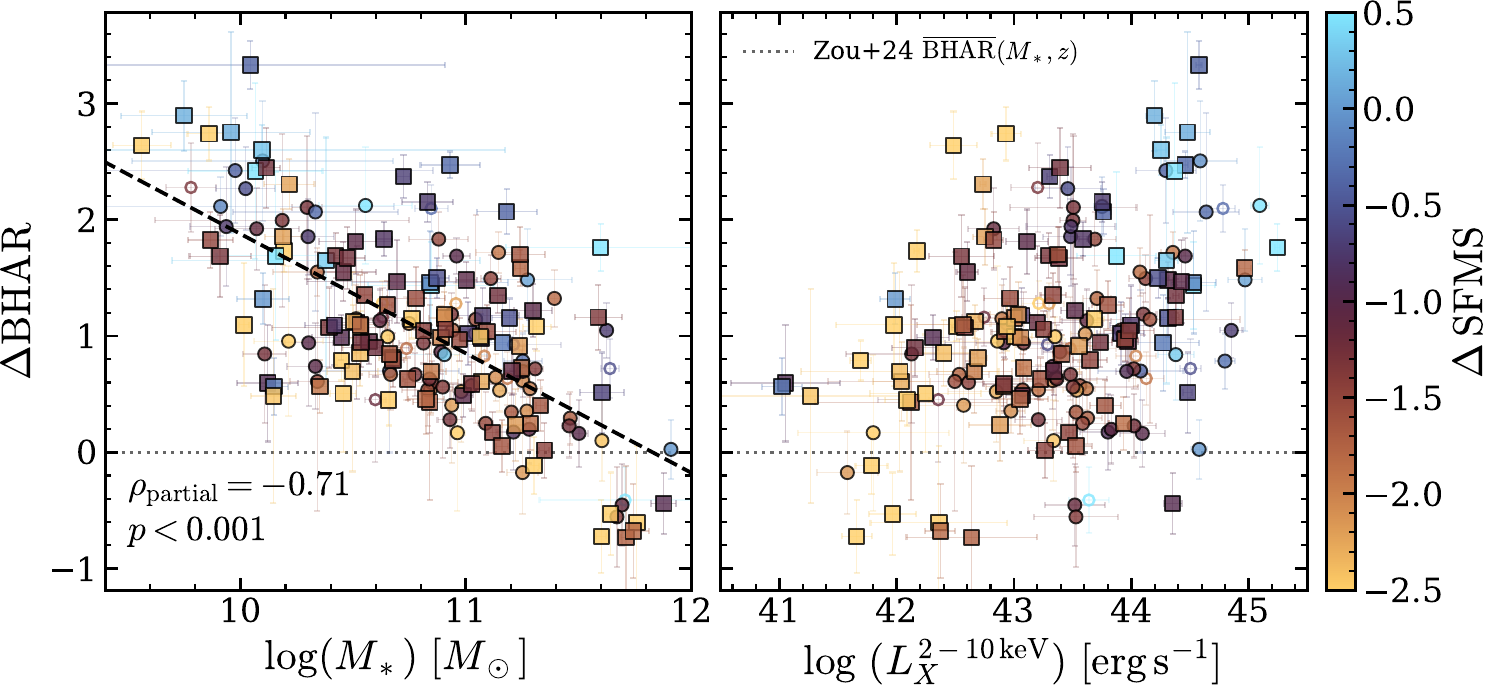}
    \caption{BHARs in the context of host galaxy properties. \textit{Top:}  2--10~keV X-ray luminosity ($L_X^{2-10\rm ~keV}$) vs.\ stellar mass ($M_*$). Filled squares, filled circles, and open circles correspond to categories A, B, and C fits, respectively. Each point's color indicates  $\Delta\mathrm{SFMS}$ according to the color bar at right. Red lines show the population-averaged $\overline{\mathrm{BHAR}}$ from \citet{Zou2024} for $z=0.3, 0.7, 1.5, 2.5$ with their associated uncertainties. \textit{Bottom left:} the BHAR offset $\Delta \mathrm{BHAR}$ (Equation~\ref{eq:bhar}) versus $M_*$. The dotted line at zero represents the galaxy's BHAR being at the population-averaged BHAR. The partial Spearman rank correlation coefficient ($\rho_{\rm partial}$) between $M_*$ and $\rm \Delta BHAR$, controlling for $L_X^{2-10\rm ~keV}$ is shown at top left: $p<0.001$ indicates the probability of obtaining the correlation under the null hypothesis of no correlation. The dashed black line shows the ordinary least-squares fit to the plotted relation. \textit{Bottom right:} $\Delta\rm BHAR$ versus $L^{\rm 2-10~keV}_X$.}
    \label{fig:BHAR}
\end{figure*}

X-ray luminosity is a probe of the instantaneous accretion rate of the central SMBH\null.
A galaxy's BHAR given the X-ray luminosity is
\begin{equation}
    \text{BHAR} = \frac{(1 - \varepsilon)L_{\rm AGN}}{\varepsilon c^2} = \frac{(1 - \varepsilon)K_X L^{2-10\mathrm{keV}}_X}{\varepsilon c^2} ~,
\end{equation}
where $\varepsilon = 0.1$ is the assumed radiative efficiency, $K_X$ is the 
bolometric correction to $L^{2-10\mathrm{keV}}_X$ from \citet{Duras2020}, and $c$ is the speed of light.

In analogy to the SFMS, the BHAR can be compared against the population-averaged value, $\overline{\rm BHAR}$. 
Following \citet{Zou2024}, the population-averaged $\overline{\rm BHAR}$ can be inferred from $M_*$ and $z$ given their published $\overline{\rm BHAR}$ maps in the $M_*$--$z$ parameter space.\footnote{\href{https://zenodo.org/records/10729248}{https://zenodo.org/records/10729248}} These $\overline{\rm BHAR}$ maps are produced from a single integral that converts AGN X-ray luminosity into the population-averaged BHAR. The integral from \citet[][ Eq.~1]{Zou2024} is
\begin{equation}
    \begin{split}
        \overline{\text{BHAR}}(M_*, z) &= \\
        \frac{(1 - \varepsilon)\,}{\varepsilon c^2} \int_{ \lambda_{\min}}^{+\infty} 
        & K_X(L^{2-10\mathrm{keV}}_X)\, L^{2-10\mathrm{keV}}_X
        \, p(\lambda \mid M_*, z) \, d\lambda ~,
    \end{split}
\end{equation}
where $\lambda = L^{2-10\mathrm{keV}}_X/M_*$ is a specific accretion rate proxy, and $p(\lambda\,\mid\,M_*, z)$ is the probability distribution 
of a galaxy hosting an AGN at a given $\lambda$ (\citealt{Zou2024} give full details).

The difference between a galaxy's instantaneous $\mathrm{BHAR}$ from $L_{\rm AGN}$ and the population average $\overline{\rm BHAR}(M_*,z)$ is then the BHAR offset,
\begin{equation}
    \Delta \mathrm{BHAR} = \log[ \mathrm{BHAR}({L_{\rm AGN}})] - \log[ \overline{\rm BHAR}(M_*,z)] ~.
    \label{eq:bhar}
\end{equation}
BHARs were converted from units of \Msol~yr$^{-1}$ to luminosities in ergs~s$^{-1}$ assuming a rest-mass energy conversion of $L_{\rm AGN}=\frac{\varepsilon}{1-\varepsilon}\,\dot{M}_{\rm BH}c^2$ in order to directly compare BHAR energy output with the inferred AGN luminosities.

These BHAR offsets are shown in Figure~\ref{fig:BHAR}. The population averaged BHARs at different redshifts are shown against the X-ray sample in the $L_X^{2-10~\rm keV}-M_*$ plane to provide context for this procedure. Most objects accrete above the shown $\rm\overline{BHAR}s$, though the bottom two plots in Figure~\ref{fig:BHAR} reveal how $\rm\Delta BHAR$ relates with $L_X^{2-10~\rm keV}$ and $M_*$.

$\mathrm{BHAR}$ deviates most from the population average for the lowest-mass galaxies. At the highest masses, the BHARs are closer to the population average. These results suggest a difference in duty cycle as the less massive galaxies seem to be going through ``growth spurts'' whereas the massive galaxies reside near the population average; they are indeed accreting, but doing so steadily and closer to the timescales associated with the timescales of galaxy evolution  that $\overline{\mathrm{BHAR}}$ tracks, i.e., $\simeq$~Gyr timescales.

However, $\rm\Delta BHAR$ is loosely linked to X-ray luminosity. Highly luminous X-ray AGN are likely not sustained over long periods of time but rather are short, stochastic bursts of SMBH growth that significantly deviate from $\overline{\rm BHAR}$. In addition, the relative accretion output of different mass SMBHs undergoing unique accretion episodes is captured through this loose link.  

$\Delta\rm SFMS$ provides additional context via color in Figure~\ref{fig:BHAR}. $\Delta\rm BHAR$ versus both $M_*$ and $L_X^{2-10\mathrm{keV}}$ shows there is no correlation between   $\Delta\mathrm{SFMS}$ and $\Delta\rm BHAR$.  $L_X^{2-10\mathrm{keV}}$ continues to correlate with $\Delta\rm SFMS$ (as also shown in Figure~\ref{fig:sfms}) whereas $M_*$ strongly correlates with $\rm\Delta BHAR$ as quantified by $\rho_{\rm partial}=-0.71$, controlling for $L_X^{2-10~\rm keV}$. The star-forming properties of a galaxy are generally decoupled from $\Delta\rm BHAR$; however, the star-forming galaxies are coincident with the most efficiently accreting SMBHs, and the growing SMBHs under short-lived duty cycles reside in the lowest-mass galaxies as reflected by $\Delta\rm BHAR$. 

\subsubsection{SMBH Growth and Galaxy Growth
}
\label{s:smbh}

The sample of X-ray galaxies studied spans a wide range of $M_*$ and $L_{\rm AGN}$ over $\simeq12$~Gyr of cosmic time. Because AGN are stochastic events within a galaxy that span a large dynamic range of accretion episodes (i.e., duty cycle), it is pertinent to attempt associating SMBH accretion episodes with galaxy growth (i.e., star-formation). Since X-ray emission probes current SMBH growth,  BHARs together with their host-galaxy properties can provide  proxies for the  link between galaxies and their SMBHs.
%


\begin{figure*}
    \centering
    \hspace{0cm}
    \includegraphics[width=0.758\linewidth,clip=true,trim=0 38.5 0 0]{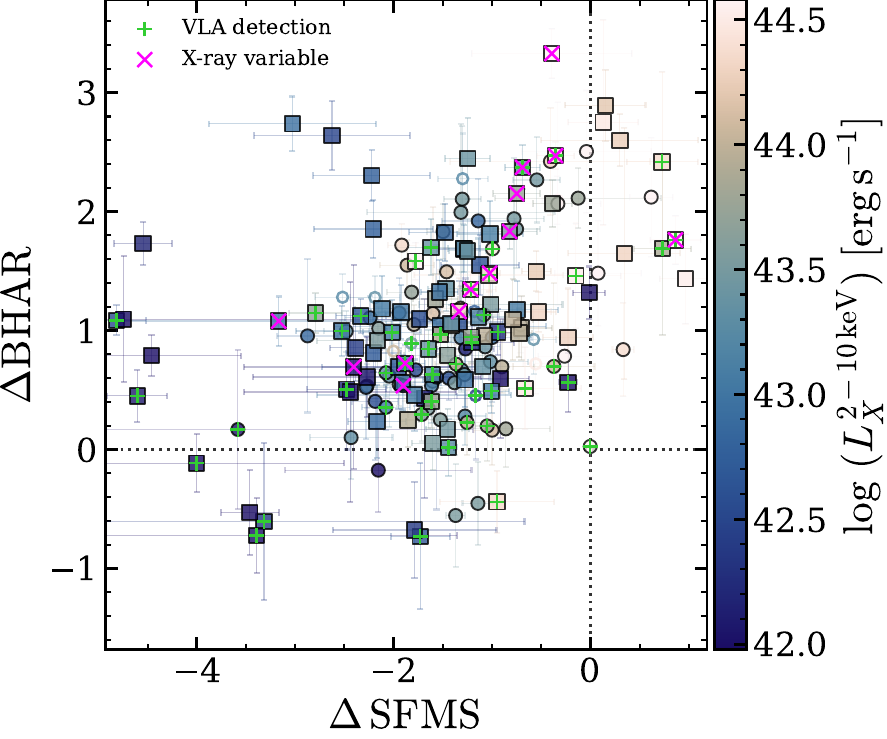}
    \includegraphics[width=0.75\linewidth]{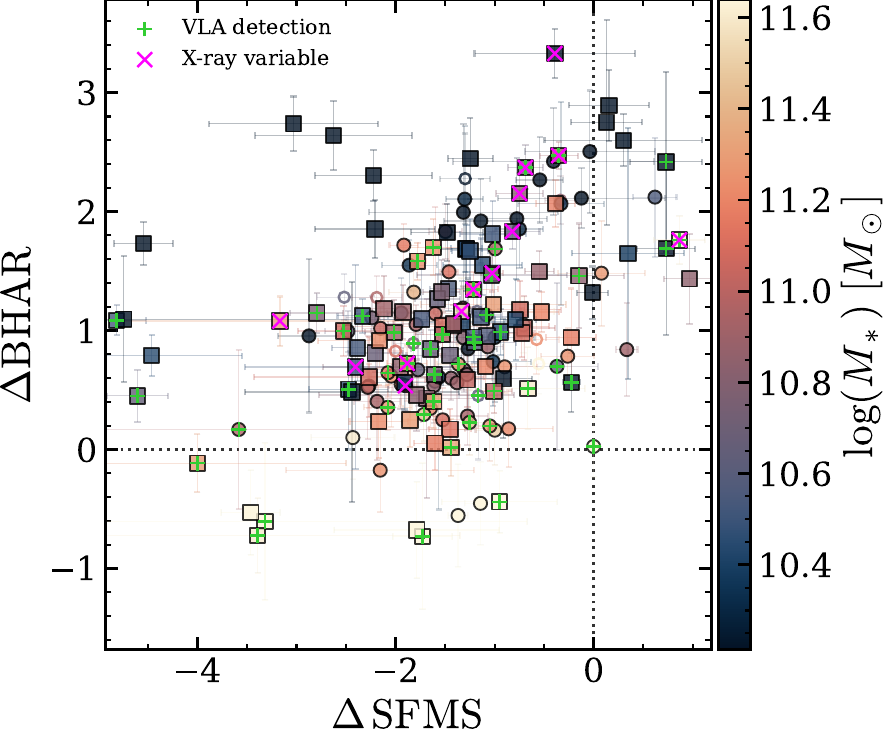}
    \caption{Black hole accretion rate offset ($\Delta\rm BHAR$) versus star-forming main sequence offset ($\Delta\rm SFMS$). \textit{Top:} Each point's color indicates $L^{\rm2-10keV}_X$ according to the color bar at right. Filled squares, filled circles, and open circles correspond to categories A, B, and C fits, respectively. Dotted lines at zero correspond to the population-averaged BHAR ($\overline{\mathrm{BHAR}}(M_*,z)$) and the SFMS. Pink crosses and green plus signs identify X-ray variable candidates and radio detections at 3~GHz, respectively. \textit{Bottom:} Same as top but with color indicating the stellar mass, $M_*$. }
    \label{fig:sfms_bhar}
\end{figure*}

\begin{figure}
    \centering
    \includegraphics[width=0.95\linewidth]{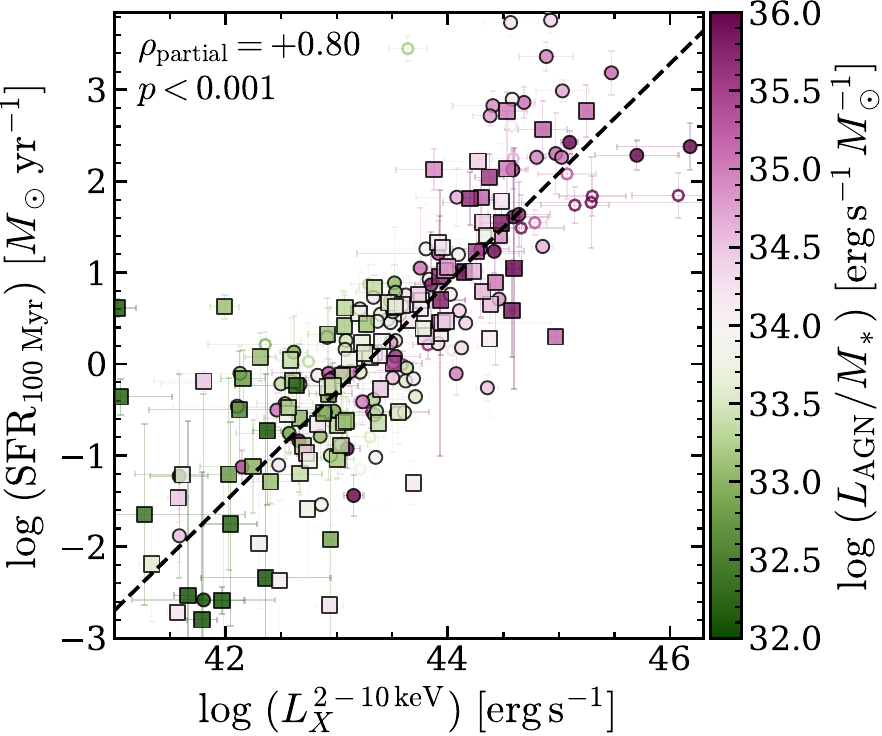}
    \caption{\texttt{CIGALE} SFR averaged over the last 100~Myr vs.\ 2--10~keV X-ray luminosity. Each point's color indicates the specific AGN luminosity $L_{\rm AGN}/M_*$ \citep{Zou2024, Aird2018} according to the color bar at right.  White points isolate the critical $L_{\rm AGN}/M_*\approx10^{34}$ range discussed in Section~\ref{s:smbh}. Filled squares, filled circles, and open circles correspond to category A, B, and C fits, respectively. The dashed black line shows the ordinary least-squares fit to the plotted log--log relation. The partial Spearman rank correlation coefficient ($\rho_{\rm partial}$) between $L_X^{2-10\rm~keV}$ and $\rm SFR_{100Myr}$, controlling for $L_{\rm AGN}/M_*$ is shown at top left: $p<0.001$ indicates the probability of obtaining the correlation under the null hypothesis of no correlation.}
    \label{fig:sfr100}
\end{figure}

Figure~\ref{fig:sfms_bhar} puts
the X-ray emission into context with the average SMBH and the SFMS\null. The less massive galaxies preferentially deviate from both $\Delta\mathrm{BHAR}=0$ and $\Delta\mathrm{SFMS}=0$, whereas more massive galaxies tend to reside at the population average for both $\Delta\mathrm{BHAR}$ and $\Delta\mathrm{SFMS}$. The most X-ray luminous sources exist near the SFMS and with heightened BHARs against the population average. In a very broad sense, this parameter space isolates distinct AGN types and phases; the most efficiently accreting SMBH in shorter-timescale accretion episodes (the upper part of the figure), the star-forming galaxies with concurrent SMBH accretion (the upper right), the X-ray emitting SMBHs that exist in longer duty cycles and lower X-ray luminosities (the bottom half of the figure), and those quenched galaxies with AGN activity (the left half of the figure). Though all X-ray sources here 
are AGN in the canonical sense  (i.e., they were luminous extragalactic X-ray detections and thus have a centrally accreting SMBH), the plots clarify how much these SMBHs are accreting relative to the wider population. Importantly, Figure~\ref{fig:sfms_bhar} reveals the instantaneous link between SMBHs and their host galaxies; most X-ray AGN reside in non-star-forming galaxies, though now the systems where AGN might promote or stunt star-formation are identifiable given this $\Delta\rm SFMS-\Delta\rm BHAR$ contextualization. 

To investigate simultaneous SMBH accretion and star-formation---especially for the most luminous X-ray sources---Figure~\ref{fig:sfr100} shows the relation for the average SFR in the last 100~Myr (SFR$_{\rm 100Myr}$). A positive correlation exists between $L_X^{2-10\rm keV}$ and SFR$_{\rm 100~Myr}$, with a partial Spearman correlation of $\rho=+0.80$, controlling for $L_{\rm AGN}/M_*$. This is particularly evident in the colormap of $L_{\rm AGN}/M_*$, which serves as a proxy for the Eddington limit and the most efficiently accreting SMBHs \citep{Aird2018, Zou2024}. At $L_X^{2-10\rm keV}\gtrsim10^{44}\mathrm{~erg~s^{-1}}$ and $L_{\rm AGN}/M\gtrsim10^{34}\mathrm{~erg~s^{-1}~M_\odot^{-1}}$, there appears to be a  narrower range in $L_X^{2-10\rm keV}$ for a given SFR\null. This suggests that the most star-forming AGN hosts are coincident with the most X-ray luminous AGN. 
This critical luminosity at $L_{\rm AGN}/M_*\simeq10^{34}~\rm erg~s^{-1}M_\odot^{-1}$ is consistent with the break in the double power law that defines the range of SMBHs that most efficiently accrete \citep{Aird2018, Zou2024}.

\begin{figure*}
    \centering
    \hspace{-0.1cm}
    \includegraphics[width=0.999\linewidth]{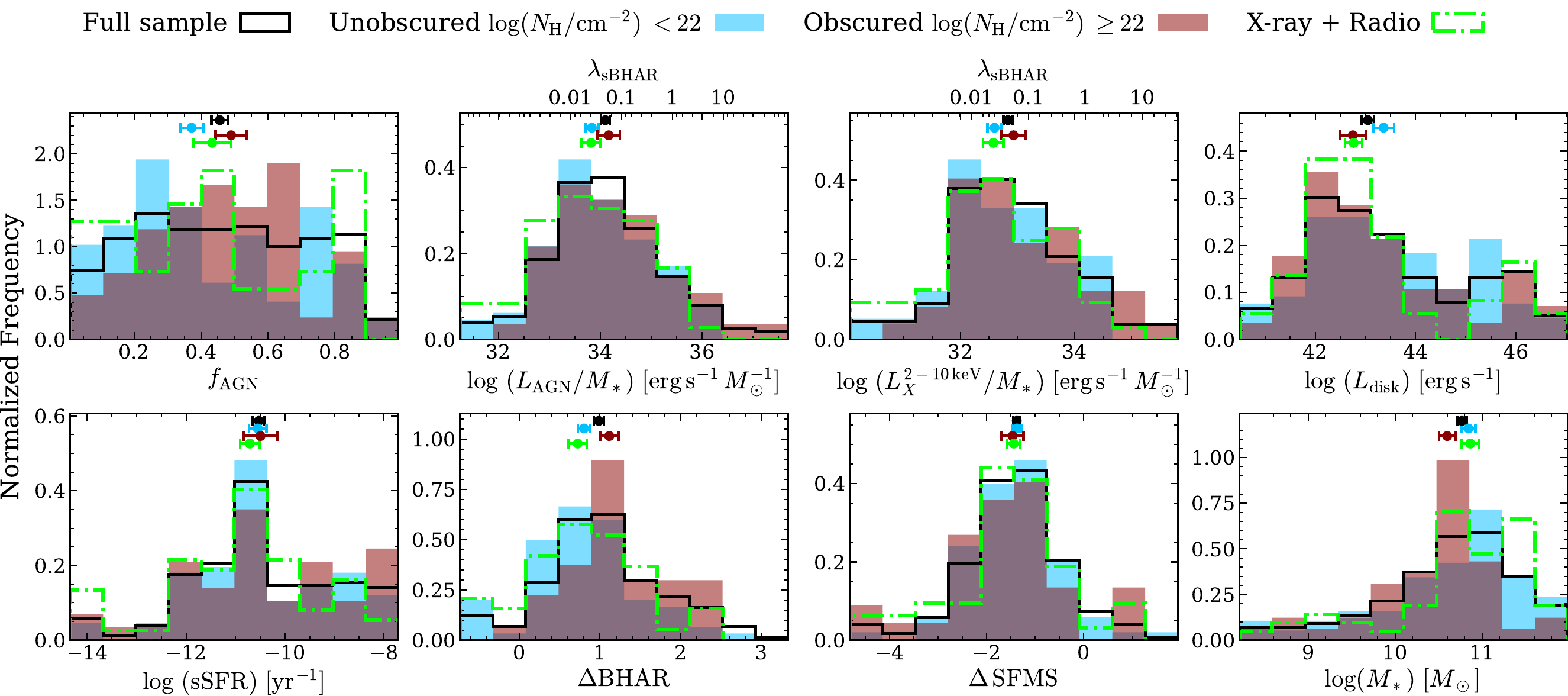}
    \caption{Histograms of AGN and galaxy properties from \texttt{CIGALE}. Each panel shows four histograms corresponding to unobscured ($\log(N_H/\rm cm^{-2})<22$) and obscured ($\log(N_H/\rm cm^{-2})>22$) X-ray sources (blue and red, respectively), galaxies with VLA counterparts (green dot-dash), and the full sample (solid black). 
    Points with horizontal error bars show the medians and standard uncertainty of the median of each property as color coded. The $\lambda_{\rm sBHAR}$ specific accretion rate proxy conversion is shown at the top axes for the second and third histograms following \citet{Aird2018, Alexander2025}.}
    \label{fig:agnHist}
\end{figure*}

\section{Discussion}
\label{sec:discussion}

In this work, the multiwavelength counterpart SEDs of the \Zhao\ and \Silver\ X-ray samples in the NEP TDF are fit with \texttt{CIGALE} to better understand AGN emission within the broader context of their host galaxies' star-formation and assembly.  Including the X-ray flux in the SED provides a powerful constraint on AGN emission and mitigates bias on SFR inferences. Given the wavelength range of observations, the results focus on using $M_*$, SFR, $L_{\rm AGN}$, $\Delta\mathrm{SFMS}$, and $\Delta\mathrm{BHAR}$ to continue disentangling the role of radiatively efficient AGN within the evolution of their host galaxies. In this section, we interpret the main empirical trends in light of existing population studies and frameworks for AGN duty cycles and coevolution with their host galaxies.

\subsection{Traditional AGN Distinctions and Properties}

The range of galaxy properties for these X-ray AGN in the sample helps evolve the picture in which SMBHs and galaxies co-evolve over cosmological timescales \citep{Kormendy2013, Madau2014, Heckman2014}. The redshift distribution peaks around $z\simeq1$, consistent with the known redshift density for X-ray AGN \citep{Ueda2014, Aird2015}. 
On cosmic scales, the BHARD and SFRD suggest that over long times and large samples galaxies and SMBHs accrete and grow in tandem. Both BHARD and SFRD peak around Cosmic Noon at $z\simeq2$ \citep{Madau2014, DSilva2025}, reinforcing the idea that SMBH growth traces the buildup of stellar mass and vice versa. 
X-ray surveys are critical for synthesizing this apparently aggregate coevolution by sampling the shorter-timescale bursts of SMBH growth associated with radiatively efficient SMBHs across a wide range of luminosities, cosmic time, and galaxy types. Different types or phases of AGN (i.e., obscured, unobscured, jetted, quasar, etc.)\ exhibit redshift evolution \citep{Aird2015, Padovani2017, Hickox_2018}, though it is still debated when and how AGN promote or stunt galaxy growth as a galaxy evolves and as the universe evolves. More importantly, it is necessary to identify the galaxy systems that host these unique AGN types and phases for detailed analysis.

In general, this sample exhibits distinct  connections between galaxies and their central SMBHs. This is best revealed by situating these AGN in the context of their host galaxy properties.  Figure~\ref{fig:agnHist} shows the bulk properties for the sample. The histograms  correspond to the X-ray unobscured ($\log(N_H/\rm cm^{-2}) < 22$) and obscured populations ($\log(N_H/\rm cm^{-2}) < 24$), the X-ray sources with VLA 3~GHz counterparts, and the full sample. 
No stark differences appear between the various AGN populations as also shown by the population medians in each panel of Figure~\ref{fig:agnHist}. It is common to distinguish AGN by identification method or AGN type (e.g., Type I vs. Type II, unobscured vs obscured). Figure~\ref{fig:agnHist} shows minimal distinction between these AGN classifications within the sample, and struggles to disentangle the nuanced relationship between galaxies and their SMBHs better revealed by SFMS and BHAR framework. The X-ray perspective and the associated context of their host galaxies is critical toward identifying the instantaneous link SMBHs and their galaxy's coevolution. 

Still, the comparisons in Figure~\ref{fig:agnHist} capture a wide dynamic range in SMBH duty cycle, accretion type, and energy output. The medians for these different classifications capture the sample majority, which are accreting SMBHs in non-star-forming galaxies. This is consistent with the earlier analysis; though, the Eddington ratio proxy, $\lambda_{\rm sBHAR}$, is shown in the second and third plots of Figure~\ref{fig:agnHist} and captures this dynamic range in the histograms, identifying the rare ``tail-ends'' of the sample demographic. The most specifically luminous AGN with $\lambda_{\rm sBHAR}\simeq1$ simply cannot maintain such accretion for long timescales, hence there are few detections of that sort. On the contrary, there is a steep ``cutoff'' at $\lambda_{\rm sBHAR}<<0.01$, likely representing the accreting SMBHs below the detection limit of this survey, or the dormant SMBHs \citep{Aird2018, Yang2018}. While the median characterizes the general X-ray AGN sample, the dynamic range of the distributions captures the timescale physics.

Interestingly, there is no immediate distinction between X-ray obscured and unobscured galaxies captured via $N_{\rm H}$. We explored if $N_{\rm H}$ showed any relationship with the various host galaxy properties already studied, but there were no clear correlations. This might suggest that nuclear gas and dust densities are largely decoupled from host galaxy activity at kiloparsec scales. Comparing this nuclear obscuration in the X-ray with obscuration of the AGN in the ultraviolet-visible can help unveil if this is the case, and we defer this to future work. This can be especially powerful out to high redshift with additional space-based imaging with HST and JWST. It is also possible that these $N_{\rm H}$ trends might have already been washed out due to the sample selection requiring a bright optical counterpart. It could be that $N_{\rm H}$ does trend with host-galaxy properties and this sample is biased against recovering that effect since we don't sample the most obscured galaxies and AGN. Future work that leverages the deep and hard X-ray data in the NEP TDF can continue to reveal known $N_{\rm H}$ relationships with galaxy morphology or mid-infrared data, and future monitoring of the NEP TDF with JWST/NIRCam and JWST/MIRI would enable this future work.

\subsection{Characterizing the Instantaneous Coevolution of Galaxies and their SMBHs}

Thanks to population-averaged characterizations like the SFMS and $\overline{\rm BHAR}$, these SMBHs emissions have been shown to exist within a diverse range of host galaxies, disentangling the more instantaneous link between galaxies and their embedded AGN. Previously, samples of large X-ray surveys have only used either the SFMS or BHARs to explore the coevolution between host galaxies and their AGN \citep{Suh2019, Leja2022, Yang2018, Zou2024}; here, we combine them into a single parameter space to provide critical perspective on the general AGN population, along with the rare relationships between SMBHs and their galaxies.

\subsubsection{The $\Delta\rm SFMS$--$\Delta\rm BHAR$ Contextualization}

The full ensemble of host galaxy and AGN properties inferred from the \texttt{CIGALE} SED fitting is shown in Figure~\ref{fig:sfms_bhar}, where SFMS and BHAR offsets give key context to these AGN. This parameterization of SED inferences helps distinguish many AGN types and phases, as listed in \S~\ref{s:smbh}. In contrast to traditional AGN distinctions based on identification method or observational effects (see Figure~\ref{fig:agnHist}), this approach understands these accreting SMBHs as \textit{events} within the life-cycle of their host galaxies. This nuance reveals the general timescales and roles of the AGN in the galaxy's evolution, in contrast to classifications that struggle to be linked together via an evolutionary sequence.

Though there is likely not one common evolutionary sequence shared by all AGN, this characterization finds that low-mass galaxies have growing SMBHs that boast more energetic accretion compared to the population-averaged BHAR for galaxies of the same mass and redshift (Figure~\ref{fig:BHAR}). In addition, these low-mass galaxies also preferentially exist at the SFMS (Figure~\ref{fig:sfms}). Though this sample is only sensitive to the lower X-ray luminosities out to $z\sim 1$ (see Figure~\ref{fig:LxVSz}), these results suggest that the growing SMBHs are among the low-mass end of the SMBH population \citep{HeckmanBest2012}, and that they are likely in ``growth spurt'' episodes of short-timescale accretion, often coincident with ongoing star-formation in the host galaxy. It is unclear if this gas-fueling is simultaneous because of shared gas reservoirs, or because of positive feedback (i.e., AGN feedback inducing star-formation); though, what is clear is the rare yet simultaneous connection between these two events in a galaxy. 

In this same vein, the characterizations with the SFMS and BHARs also reveal a clear link between star-forming AGN hosts and their X-ray luminosities. The causal link between star-formation and X-ray luminosity is still debated \citep{Leja2022, Cristello2024, Suh2019, Mountrichas2022}, but is has become clear that there exist a population of radiatively efficient SMBHs accreting at the highest-luminosities that are coincident with star-formation \citep{Alexander2025, Kirkpatrick2020}. This work finds that there is a relationship between SFR and $L_X$ (Figure~\ref{fig:sfr100}), though higher X-ray luminosities don't necessarily promise higher SFRs \citep{Alexander2025}. However, it is true that if a galaxy is star-forming and hosts an X-ray emitting SMBH, the SMBH is most often in a radiative-mode of accretion that boasts exceptional X-ray luminosities. 

The relationship between star-formation and SMBH accretion remains complex. The computed correlation between $\rm SFR_{100~Myr}$ and $L_X^{2-10\rm~keV}$ in Figure~\ref{fig:sfr100} ($\rho_{\rm partial}=+0.80$) suggests both SMBH accretion and star-formation draw from the same gas reservoirs over similar timescales in these systems. This timescale of star-formation is rare to catch when imposed against galaxy evolution on the order of billions of years, so it is indeed rare to see $L_X$ always correlate with star-formation. This might also be because the most luminous X-ray sources do not always correlate with starburst activity, though it might be that the most starburst-type AGN hosts do correlate with the most radiatively efficient AGN. This picture is supported by \citet{Aird2019}, who showed that for main-sequence galaxies, the average AGN incidence and specific accretion rates increase linearly with SFR, but individual systems scatter widely in $L_X$\null. \citet{Alexander2025} supported this interpretation and argued that over evolutionary timescales in galaxies, multiple AGN events may occur when the SMBH accretes and grows. This is evident in the broad distribution of specific BHARs (which serve as a proxy for Eddington luminosity) across a narrow range of sSFR (see Figure~\ref{fig:agnHist}). 
In contrast, selecting galaxies of a given SFR and averaging $L_X$ reveals this observed correlation \citep{Alexander2025, Dai2018}.

\subsubsection{AGN Feedback}

In the present sample, the host galaxies of  X-ray AGN generally fall below the SFMS\null. Some are defined as quenched, consistent with the need for AGN quenching in massive galaxies to match the stellar mass function---especially in simulations \citep{Croton_2006, Somerville2015, Pillepich2018}. Most of the sample likely exists below the SFMS because star-formation is rare and episodic within a galaxy, hence the observed effect is reflective of AGN existing in galaxies but not largely influencing it via feedback throughout most of the galaxy's life. However, the subset of quenched galaxies defined by the SFMS spans the full range in stellar mass. Though this analysis does not claim the role of AGN in quenching the galaxies---nor is there any sense of ``direction'' in which a galaxy is moving in $\Delta\rm SFMS$--space (i.e., toward or away from the SFMS)---, what can be said is there are quenched galaxies with AGN presence. The low-mass galaxies that are defined as quenched might have suffered stunted galaxy growth from AGN feedback, though for the high-mass galaxies it is less clear if quenching comes from the AGN or general stellar mass build up. 

The VLA 3~GHz imaging and analysis by \citet{Hyun2023, Willner2026} revealed a source with offset radio emission, likely coming from a radio lobe. This source is VLA ID~194 = XMM ID~222 in this work. This source is one of the most interesting sources in the field because it is also suggested as Compton thick \citep{Creech2025} and shows signs of $N_{\rm H}$ variability (Creech et al.\ in prep.). The offset radio emission could indeed influence the evolution of this Seyfert-like galaxy, which in this work does not have a high SFR\null. Given the high $N_{\rm H}$ and low SFR with possible extended radio structure offset from the galaxy center, this source likely has an AGN with a longer duty cycle and serves as an interesting candidate for spectroscopic analysis in the visible and near-infrared to identify the effect of outflows or other AGN feedback on  the host galaxy. The rest of the radio imaging suggests that most radio sources have central emission coming from a point source, thus suggesting minimal jetted structure in the present X-ray AGN sample. A clear double-lobed radio source in the VLA imaging is VLA ID~141, though this source is not matched to an XMM source. XMM ID~224 is a counterpart to VLA ID~201, which is suggestive of double-lobed structure, and the JWST/NIRCam imaging suggests merger-induced tidal features. 

Currently, the field has spectra from MMT/Binospec and MMT/Hectospec. Future work could benefit from using the identified starburst and quenched candidates in the sample and model their optical spectra to identify outflows and obscuration of emission lines in an attempt to probe the kinematics of these AGN feedback and fueling mechanisms. 

\subsubsection{Coincident Accretion}

There is a small population of highly luminous X-ray sources with coincident star-formation, consistent with previous work that finds X-ray AGN can occupy the SFMS at increased $L_X$ \citep{Mountrichas2022, Cristello2024}. However, recent analyses using large X-ray samples also contend that, at fixed $M_*$, the typical SFR of AGN hosts tracks that of non-AGN hosts, with subtle differences that depend on $L_X$, environment, or selection method \citep{Suh2019, Leja2022, Mountrichas2023}. 

This analysis reports a strong correlation between the offset from the SFMS and the specific AGN luminosity (Figure~\ref{fig:sfms}), reflected by $\rho_{\rm partial}=+0.73$. Debate around X-ray AGN and their relationship to the SFMS revolves around $L_X$ and $M_*$; however, for individual galaxies, $L_X$ and $M_*$ can range by orders of magnitude, thus making it difficult to find common ground in SFMS studies. This problem is especially dependent on the different SFH prescriptions and SED analyses done through many public codes, e.g., \texttt{CIGALE}.

Using the specific AGN luminosity, $L_{\rm AGN}/M_*$, mitigates distance dependence and selection effects while also preserving the relative degree of accretion of the central SMBH with respect to the host galaxy's maturity (i.e., stellar mass build up, $M_*$). Thus, Figure~\ref{fig:sfms} highlights that the most radiatively efficient AGN coincide with star-forming galaxies. This complements the positive correlation with the 100~Myr averaged SFRs and X-ray luminosity in Figure~\ref{fig:sfr100}, continuing to suggest that galaxies and their central SMBHs accrete and grow together in dramatic fashion. This simultaneous fueling of the central SMBH and star-formation is common among the low-mass galaxies in the sample, though it is not exclusive to higher-mass galaxies. Here, the picture is that the growing SMBH population---especially among less mature galaxies---feeds off pristine gas reservoirs that simultaneously fuel star-formation and mass build up of the galaxy. For more mature galaxies with the higher stellar masses ($\log(M_*) \gtrsim 11$), merger-induced AGN and starbursts are a likely ignition mechanism. 

Not only is the positive correlation between $\Delta\rm SFMS$ and $L_{\rm AGN}/M_*$ clear, but also are the AGN duty cycles involved. Most of the sample exists below the SFMS and below the characteristic specific AGN luminosity that represents the break in the double power law of Eddington luminosities for X-ray AGN, $L_{\rm AGN}/M_* < 10^{34} \rm ~erg~s^{-1}~M_\odot^{-1}$ \citep{Aird2018, Zou2024}. This result suggests that most accreting SMBHs exist in moderate accretion modes that track timescales longer than star-formation episodes in galaxies, yet the modes of more efficient accretion that approach the Eddington luminosity are short-lived and consistent with timescales associated with the fueling of star-formation.

\subsubsection{Duty Cycles}

Converting $L_X$ into BHARs following \citet{Yang2018, Zou2024} captures the vast range of timescales associated with SMBH accretion within the evolution of their galaxies. 
\citet{Yang2018} showed that $\overline{\rm BHAR}$ depends on $M_*$ and $z$ and approximately tracks the SFRD evolution. This suggested that more massive galaxies host more rapidly growing SMBHs earlier on in the Universe's history. \citet{Aird2018} took the largest sample of Chandra sources and measured the full probability distribution of specific accretion rates as a function of $M_*$ and $z$, finding a broad and  roughly power-law distribution of specific accretion rates with a cutoff near the Eddington limit.

Within this framework, the significant deviations in $\Delta\rm BHAR$ are a direct manifestation of the stochasticity associated with AGN accretion. One can interpret the offsets as a proxy for duty cycle, with the largest offsets corresponding to the shortest-lived SMBH growth episodes that drive rapid increases in $M_{\rm BH}$. This is consistent with results suggesting that the more massive end of the SMBH mass function does not grow, and that most SMBH growth occurs in lower-mass SMBHs \citep{HeckmanBest2012}. Finding that the high-mass galaxies' SMBHs have BHARs consistent with the population average $\overline{\rm BHAR}$ is in line with \citet{Yang2018, Zou2024}, who infer that massive systems maintain a more ``steady'' average accretion, while low-mass systems are preferentially detected in episodic SMBH growth. 

Duty cycles provide the context for the large spread in $\Delta \rm BHAR$, $\Delta\rm SFMS$, and $L_X$\null. Of course, not all AGN ignite and accrete in the same way; however, this interpretation connects the many types of accreting SMBHs within their diverse host galaxy environments in order to identify the general properties of X-ray detected AGN, as well as the rare systems sporting the short-timescale and  instantaneous coevolution between a galaxy and the central SMBH.

\subsection{The NEP TDF Landscape}

\subsubsection{X-ray Variable Sources}
Ongoing work by Creech et al.\ (ApJ, submitted)\ (C26) and Silver et al.\ (ApJ, submitted)  (S26b) has identified X-ray variable sources in then NEP TDF via spectral and photometric variability, respectively. Photometrically variable sources in the XMM-Newton soft X-ray bands are 111/213/222/299/305/318/327/329/356/375/422. IDs 131/222/299 are variable in the NuSTAR hard X-ray bands. The C26 analysis found that IDs 108/111/161, 222/277/286/299/356 suggest variability in their X-ray spectra. 

Both C26 and S26b suggested that IDs 222/299 are among the most interesting and secure X-ray variable sources. ID 222 falls within the JWST/NIRCam footprint and boasts a central point-like signature at 4.4~\micron, distinct from the resolved spiral structure of the host galaxy\citep{Ortiz2024}. This source suggests hydrogen column density ($N_{\rm H}$) variability, and it appears to have transitioned from the Compton-thick regime during the multi-year monitoring campaign. 

ID 299 is the known blazar and brightest X-ray source in the field. ID~299 is also a strong ``cold quasar'' candidate, following the presentation by \citet{Kirkpatrick2020} of highly luminous, unobscured, and starbursting quasars. The source has a $\rm SFR = 501~M_\odot~yr^{-1}$ and an $L_{\rm AGN}=46.8~\rm erg~s^{-1}$. Other strong candidates for these cold quasars include IDs 35/71/347 (all Category A), whereas the other  
sources are Category B and may be more spurious SED inferences.

From Figure~\ref{fig:sfms_bhar}, the X-ray variable candidates tend to be among the more luminous X-ray sources, likely assisting their detections as variable over the multi-year  observations in the NEP TDF\null. In addition, their SFRs are near the SFMS, supporting the variable classification if there is indeed inflowing gas that could be related to episodic star-formation and stochastic gas accretion. Some of the X-ray variable sources are also radio detections at 3~GHz. These radio detections could be from star-formation, further suggesting the stochasticities involved in radiatively efficient X-ray AGN and episodic star-formation, making these X-ray variable candidates prime targets for future monitoring and spectroscopic follow-up in order to identify inflows or outflows of  possible gas accretion on parsec to kiloparsec scales.

\subsubsection{Limitations and Next Steps}

The NEP TDF is rich in wavelength coverage and depth, thus enabling this work and the results. Given the SED inferences made for these X-ray sources, there exist prime candidates for future work. Many parts of the NEP TDF also stand to greatly benefit from continued wavelength coverage and photometric monitoring. About $1/3$ of the sample suffers from less reliable photometric redshift solutions, and improved photometry in this field will better enable complementary analyses of these X-ray sources across all wavelengths.

These AGN, along with the deep- and space-based imaging, provide the ability to further understand the role of morphology with AGN incidence. Recent work with JWST/NIRCam by \citet{Bonaventura2025, Bonaventura2026} studied a large sample of infrared-selected AGN out to Cosmic Noon, parameterized the morphologies of these AGN hosts, and found a substantial preference for merger-induced morphologies among mid-infrared selected AGN. These merger-induced AGN suggest an evolutionary pathway between obscured and unobscured AGN through a blowout phase, similar to the radiation-regulated unification scenario \citep{Almeida2017, Ricci2022}. The NEP TDF has  the deepest hard X-ray data to date and a unique set of ancillary multiwavelength observations, though the lack of mid-infrared coverage and JWST/NIRCam coverage for the full X-ray population prohibits an analysis parallel to this recent work. In any case, the qualitative assessments from Figure~\ref{fig:RGB} do show some merger-induced morphologies (i.e., tidal tails, galaxy groups, irregular shapes), though other X-ray AGN appear secular and disc-like. Further structural analysis of the full X-ray sample in the NEP TDF would greatly benefit from continued monitoring with JWST/NIRCam and JWST/MIRI to complement the deep and multi-epoch X-ray data in the hard and soft bands out to high redshift. 

This work also identifies priority, high-interest candidates for future study and follow-up. The systems with coincident AGN and star-forming activity warrant spectroscopic follow-up to disentangle the gas-fueling mechanisms at work in these systems around Cosmic Noon. Further, the sample's multi-epoch X-ray coverage can be greatly complemented by continued monitoring of the NEP TDF via new JWST/NIRCam observations, which can enable time-domain coverage out to high redshift, complementing the existing optically variable sources presented by \citet{Obrien2024} with HST.

\section{Conclusion}

Available multiwavelength data for host-galaxies of the \Zhao\ and \Silver\ X-ray samples
show that most sampled AGN reside in host 
galaxies with minimal star-formation for their mass. The selected of X-ray AGN in the NEP TDF results in 261 sources with $\geq4$ ultraviolet, visible, and infrared broadband photometry giving: 127 high-fidelity category A, 115 fair category B, and 19 flagged category C SED fits.
Though the selected AGN hosts are not actively star-forming, there are ``outlier'' X-ray sources that exist within star-forming, quenched, and gas-rich systems. This is best encapsulated in the $\Delta\rm SFMS$--$\Delta\rm BHAR$ parameter space, which pinpoints the common galaxy type of X-ray selected AGN (which is largely an effect of observing short-timescale events like SMBH accretion or star-formation superimposed on galaxy evolution across billions of years) against rarer AGN--galaxy relationships that might promote or stunt galaxy growth. The highest-SFR host galaxies  preferentially coincide with heightened AGN luminosities, suggesting simultaneous gas fueling and short bouts of accretion that likely contribute to significant mass buildup. This relationship is strongly suggested in Figure~\ref{fig:sfms} and Figure~\ref{fig:sfr100}, where X-ray luminosity and SFR suggest a tight positive correlation, strongest with respect to specific AGN luminosity. These host-galaxy properties are pivotal for understanding the X-ray AGN population's place in galaxy evolution because they identify the critical galaxies that stand out from the typical framework of SMBH--galaxy coevolution. These include the ``cold quasar'' picture of starburst--AGN activity, low-mass galaxies with SMBH
``growth spurts'' characteristic of the growing SMBH population,  high-mass systems 
reflecting steady ``maintenance mode'' accretion similar to the massive end of the SMBH mass 
function, and quenched systems with AGN, which is likely the quenching agent in the lowest mass galaxies. 

In particular, we report:
\begin{itemize}

    \item \textbf{SFR--$\boldsymbol{L_X}$ correlation:} Where a galaxy exists on the SFMS is strongly correlated to the specific AGN luminosity, $L_{\rm AGN}/M_*$, verified by a partial Spearman rank correlation coefficient of $+0.73$ after controlling for $M_*$. In addition, the 100~Myr averaged star-formation and X-ray luminosity are positively correlated with partial Spearman rank correlation coefficient of $+0.80$ after controlling for mass dependence. This result suggests that the star-forming population of radiatively accreting SMBHs are fueled from the same gas reservoirs, accrete on the same timescales, and represent a critical AGN phase that results in significant mass build up and gas inflow within a galaxy. 

    \item \textbf{Cold Quasars:} There exists a critical $L_X\simeq10^{44}~\rm erg~s^{-1}$ and specific AGN luminosity $L_{\rm AGN}/M_*\simeq10^{34}~\rm erg~s^{-1}$ in which star-forming galaxies are coincident with the most luminous SMBH accretion.  These systems are similar to cold quasars \citep{Kirkpatrick2020}. The strongest cold quasar candidates  are IDs 35/71/299/347. In addition, a majority of the X-ray variable candidates exist near the SFMS, suggestive of stochastic bouts of accretion unto the SMBH and across the kiloparsec scales of the host galaxy. Future spectroscopic analysis can disentangle the kinematics and fueling mechanisms in these systems.

    \item \textbf{$\boldsymbol{\Delta\mathrm{SFMS}$--$\Delta\mathrm{BHAR}}$ framework:} Most radiative-mode AGN reside within non-star-forming galaxies. However, the X-ray perspective captures an array of accretion timescales fueling both the SMBH and the star-formation of the galaxy. This framework  provides powerful context to the instantaneous relationship between SMBHs and their host galaxies. Traditional AGN distinctions (i.e., obscured, unobscured, radio) do not reveal the same distinctions in galaxy demographics (Figure~\ref{fig:agnHist}),  highlighting the need to situate X-ray AGN within the context of their host galaxies in order to draw meaningful conclusions about the instantaneous link between galaxies and their central SMBHs.


\end{itemize}

\begin{acknowledgments}

ROIII dedicates this work to Jesus Christ. ROIII acknowledges that this work is based upon support by NASA under award number 80GSFC24M0006, cooperative under the Southeastern Universities Research Association (SURA) subaward number 143919-Z6630204 to the University of Maryland, College Park in the Center for Research and Exploration in Space Science and Technology (CRESST) II project funded by NASA. 

This work is based on observations associated with programs JWST-GTO-2738 and 1176 made with the NASA/ESA/CSA James Webb Space Telescope (JWST) and observations associated with programs HST-GO-15278, 16252 and 16793 made with the NASA/ESA Hubble Space Telescope (HST). 
The JWST and HST data were obtained from the Mikulski Archive for Space Telescopes at the Space Telescope Science Institute (STScI), which is operated by the Association of Universities for Research in Astronomy, Inc. (AURA), under NASA contracts NAS\,5-03127 (JWST) and NAS\,5-26555 (HST).  This work uses data obtained at the MMT Observatory, a facility jointly operated by the University of Arizona and the Smithsonian Institution.
RAW, SHC, and RAJ acknowledge support from NASA JWST Interdisciplinary Scientist grants NAG5-12460, NNX14AN10G and 80NSSC18K0200 from GSFC.
RAJ, RO, RAW, SHC, AMK, BLF and CNAW acknowledge support from HST grants HST-GO-15278.*, 16252.* and 16793.* from STScI, which is operated by AURA under contract NAS\,5-26555 from NASA.
CNAW acknowledges funding from the JWST/NIRCam contract NASS-0215 to the University of Arizona.

Data presented in this article were obtained from the Mikulski Archive for Space Telescopes (MAST) at the Space Telescope Science Institute. 
The NIRCam observations used here can be accessed via \dataset[10.17909/jtd6-af15]{http://dx.doi.org/10.17909/jtd6-af15}, the NIRISS observations via \dataset[10.17909/7xzs-bb33]{https://doi.org/10.17909/7xzs-bb33}, and the HST observations via 
\dataset[10.17909/wv13-qc14]{https://doi.org/10.17909/wv13-qc14}.

M. Mezcua acknowledges support from the Spanish Ministry of Science and Innovation through the project PID2024-159201NB-C22. This work was also partly supported by the Spanish program Unidad de Excelencia Mar\'ia de Maeztu CEX2020-001058-M, financed by MCIN/AEI/10.13039/501100011033, and by the MaX-CSIC Excellence Award MaX4-SOMMA-ICE.

This research is based [in part] on data collected at the Subaru Telescope, which is operated by the National Astronomical Observatory of Japan. We are honored and grateful for the opportunity of observing the Universe from Maunakea, which has the cultural, historical, and natural significance in Hawaii.

This research used data obtained with the Dark Energy Spectroscopic Instrument (DESI). DESI construction and operations is managed by the Lawrence Berkeley National Laboratory. This material is based upon work supported by the U.S. Department of Energy, Office of Science, Office of High-Energy Physics, under Contract No. DE–AC02–05CH11231, and by the National Energy Research Scientific Computing Center, a DOE Office of Science User Facility under the same contract. Additional support for DESI was provided by the U.S. National Science Foundation (NSF), Division of Astronomical Sciences under Contract No. AST-0950945 to the NSF’s National Optical-Infrared Astronomy Research Laboratory; the Science and Technology Facilities Council of the United Kingdom; the Gordon and Betty Moore Foundation; the Heising-Simons Foundation; the French Alternative Energies and Atomic Energy Commission (CEA); the National Council of Humanities, Science and Technology of Mexico (CONAHCYT); the Ministry of Science and Innovation of Spain (MICINN), and by the DESI Member Institutions: www.desi.lbl.gov/collaborating-institutions. The DESI collaboration is honored to be permitted to conduct scientific research on I’oligam Du’ag (Kitt Peak), a mountain with particular significance to the Tohono O’odham Nation. Any opinions, findings, and conclusions or recommendations expressed in this material are those of the author(s) and do not necessarily reflect the views of the U.S. National Science Foundation, the U.S. Department of Energy, or any of the listed funding agencies.

\end{acknowledgments}
\software{
    \texttt{Astropy} \citep{Astropy2013, Astropy2022},
    \texttt{SExtractor} \citep{sextractor},
    \texttt{CIGALE} \citep{Boquien2019, Yang2022}
}

\facilities{James Webb Space Telescope (\textit{JWST}/NIRCam/NIRISS), Hubble Space Telescope (\textit{HST}/ACS/WFC \& \textit{HST}/WFC3/UVIS), XMM-Newton, NuSTAR, JCMT-SCUBA-2, VLA, Subaru (\textit{Hyper-Suprime Cam}), WISE, MMT (\textit{Binospec, MMIRS}), DECaLS, GALEX }
\bibliography{bib}{}
\bibliographystyle{aasjournalv7}

\appendix

\section{\texttt{CIGALE} Parameters}

\begin{table*}[ht]
\centering
\small
\renewcommand{\arraystretch}{1.08}
\setlength{\tabcolsep}{3pt}

\caption{Custom SED fitting parameters for \texttt{CIGALE}. Only those that differ from the default are shown.}
\label{tab:sedparams-custom}

\begin{tabular*}{0.96\textwidth}{@{\extracolsep{\fill}} ll
  p{0.28\textwidth} p{0.11\textwidth} p{0.30\textwidth}@{}}
\hline
\textbf{Module} & \textbf{Parameter} & \textbf{Description} & \textbf{Default} & \textbf{Custom} \\
\hline

\multicolumn{5}{@{}l}{\textbf{SFH (sfhstochastic\_carvajal2025)}} \\
& tau\_main       & E-folding time (Myr)                        & 2000.0       & 500, 1500, 7500 \\
& age             & Oldest stellar age (Myr)                    & 5000         & 100, 300, 1000, 3000, 5000, 8000, 12000 \\
& tau\_break      & Decorrelation timescale (Myr)               & 150          & 100, 500 \\
& sigma           & SFR variability amplitude (dex)             & 0.4          & 0.1, 0.4 \\
& N\_SFH          & Number of SFH seeds                         & 0            & \texttt{np.arange(5)} \\
\hline

\multicolumn{5}{@{}l}{\textbf{SSP (cb19)}} \\
& imf             & Initial Mass Function                       & 0 (Salpeter) & 1 (Chabrier) \\
& metallicity     & Stellar metallicity                         & 0.02         & 0.0005, 0.02 \\
\hline

\multicolumn{5}{@{}l}{\textbf{Dust Attenuation (dustatt\_modified\_starburst)}} \\
& E\_BV\_lines    & Nebular $E(B{-}V)$                          & 0.3          & 0.1 \\
& E\_BV\_factor   & Continuum $E(B{-}V)$ factor                 & 0.44         & 0.1, 0.44 \\
& power-law\_slope & Attenuation curve power-law                & 0.0          & $-1.0$ \\
\hline

\multicolumn{5}{@{}l}{\textbf{Dust (dl2014)}} \\
& qpah            & PAH fraction                                & 2.5          & 3.5 \\
& umin            & Min.\ radiation field                       & 1.0          & 10 \\
\hline

\multicolumn{5}{@{}l}{\textbf{AGN (skirtor2016)}} \\
& t               & Optical depth at 9\,$\mu$m                  & 7            & 5, 9 \\
& oa              & Torus half-opening (deg)                    & 40           & 20 \\
& delta           & UV slope deviation from disk                & 0.0          & $-0.3$, $-0.1$ \\
& i               & Inclination (deg)                           & 30           & 10, 40, 80 \\
& fracAGN         & AGN fraction                                & 0.1          & 0.01, 0.15, 0.3, 0.5, 0.7, 0.85, 0.99 \\
& lambda\_fracAGN & AGN frac.\ $\lambda$-range ($\mu$m)         & 0/0          & 0.1/30 \\
\hline

\multicolumn{5}{@{}l}{\textbf{X-ray (lopez24)}} \\
& alpha\_irx      & $\log[\nu L_{\nu}(12\,\mu\text{m})/L_X]$    & 0.1, \ldots, 0.6 & 0.0, 0.2, 0.4, 0.6 \\
\hline

\multicolumn{5}{@{}p{0.96\textwidth}@{}}{\textbf{Redshifts} \quad
The redshift grid spans $z=0.1$--5 in $\log(1+z)$ space for 500 redshifts.} \\
\hline
\end{tabular*}

\tablecomments{The SED run with only ultraviolet, visible, and near-mid-IR photometry excludes the X-ray module, though redshift is a free parameter. The SED run with X-ray, ultraviolet, visible, and infrared fits at the best-fit redshift from the first SED run so that it is fixed.}
\end{table*}

\section{X-ray Fluxes}
XMM and NuSTAR intrinsic fluxes corrected for absorption following methods in \citet{Creech2025}. These intrinsic fluxes are then input into \texttt{CIGALE} for SED fitting.

\begin{figure*}[h]
    \centering

    \includegraphics[width=0.4\linewidth]{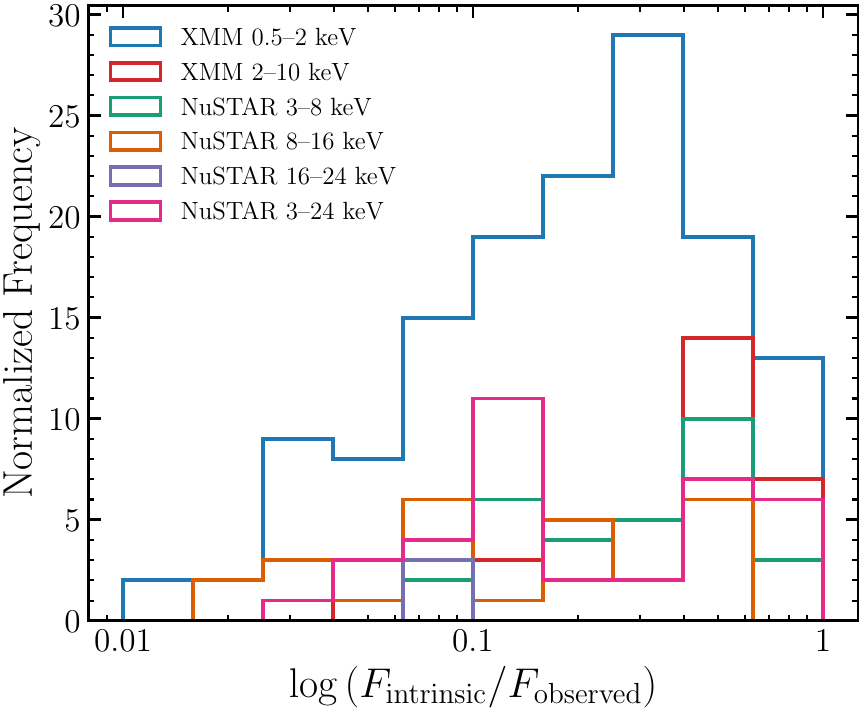}
    \caption{Histograms of the amount X-ray of obscuration. The six energy bands relevant to this work are shown as indicated in the legend.  Obscuration is given as the logarithmic ratio of observed to intrinsic X-ray flux. }
    \label{fig:intrinsic-flux}
\end{figure*}

\section{CIGALE Systematics}

For sources with photometric redshifts, we choose to assign the redshift solution based on the SED fit category (i.e., A, B, and C). See \S~\ref{sec:samplequality} for full details. Since photometric redshift solutions range from the maximum likelihood redshift ($z_{\rm ML}$), which is where the $P(z)$ curve peaks, \texttt{CIGALE} produced \texttt{best.universe.redshift}, and \texttt{bayes.universe.redshift}, we investigated any bias introduced from this procedure on the host galaxy properties $M_*$ and $L_{\rm AGN}$.  We show the difference in $L_X$ and $M_*$ evaluated at these two 
\texttt{best.universe.redshift} and \texttt{bayes.universe.redshift} redshifts. For category A, the $P(z)$ curve peaks at \texttt{bayes.universe.redshift}, and for category B, $P(z)$ generally peaks at the \texttt{best.universe.redshift}. Figure~\ref{fig:systematics} shows that category B fits might  be biased to slightly higher $L_X$, but not to a degree that ruins the integrity of the inferences and discussed results. 
\begin{figure}
    \centering
    \includegraphics[width=0.7\linewidth]{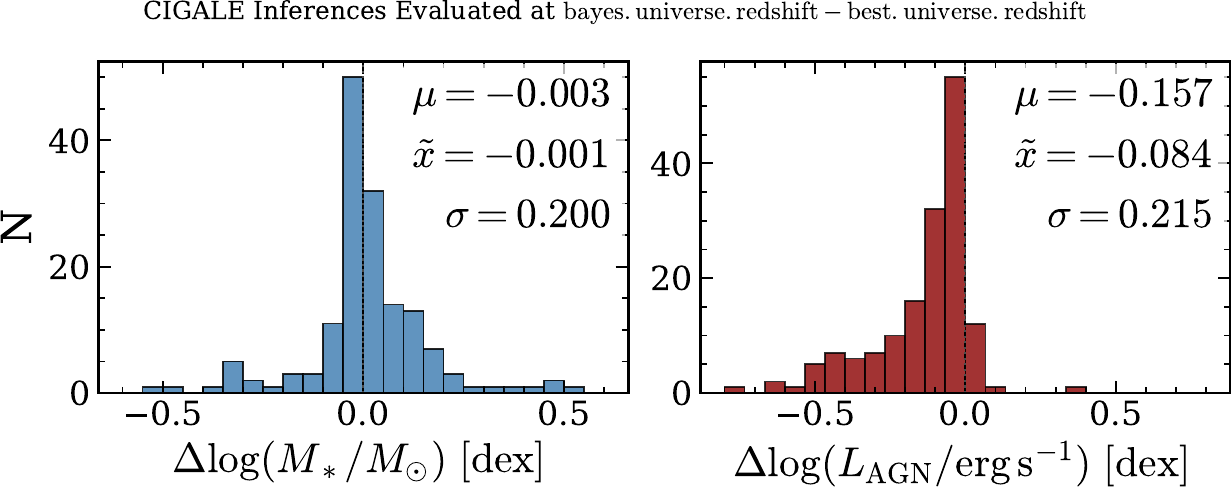}
    \caption{Difference between \texttt{CIGALE}'s ``Bayes''   
     and ``best'' outputs in $M_*$ and $L_X$ for sources with photometric redshift solutions. The mean, median, and standard deviation of the differences is shown in each panel in text.}
    \label{fig:systematics}
\end{figure}




\begin{deluxetable*}{cccccccccccccccc}
\tablecaption{Source Catalog Identifications, Positions, Quality Metrics, and Inferences for the 268 sources fit in \texttt{CIGALE}.}
\tabletypesize{\normalsize}
\label{table:z_fAGN_class}
\setlength{\tabcolsep}{2pt}
\renewcommand{\arraystretch}{0.88}
\tablewidth{0pt}
\tablehead{
  \colhead{XMM ID} & \colhead{Z24 X-ID} & \colhead{S26 X-ID} & \colhead{NuSTAR ID} & \colhead{Z24 N-ID} & \colhead{S25 N-ID} & \colhead{VLA ID} & \colhead{R.A.} & \colhead{Decl.} & \colhead{$z$} & \colhead{$Q_z$} & \colhead{$\chi^2$} & \colhead{$\chi^2_\nu$} & \colhead{$N_{\rm filt}$} & \colhead{$f_{\rm AGN}$} & \colhead{Class} \\
  \colhead{(1)} & \colhead{(2)} & \colhead{(3)} & \colhead{(4)} & \colhead{(5)} & \colhead{(6)} & \colhead{(7)} & \colhead{(8)} & \colhead{(9)} & \colhead{(10)} & \colhead{(11)} & \colhead{(12)} & \colhead{(13)} & \colhead{(14)} & \colhead{(15)} & \colhead{(16)}
}
\startdata
5 & 262 & \nodata & \nodata & \nodata & \nodata & \nodata & 260.0903320 & 65.9472660 & 0.813$^{a}$ & 0.74 & 25.17 & 2.91 & 5 & 0.19 & G \\
7 & \nodata & 202 & \nodata & \nodata & \nodata & \nodata & 260.1084590 & 65.8040540 & 4.455$^{a}$ & 4.61 & 5.28 & 0.92 & 5 & 0.62 & Q \\
10 & 52 & 69 & 99 & \nodata & 53 & \nodata & 260.1358340 & 65.8898700 & 1.422 & \nodata & 9.51 & 2.00 & 5 & 0.57 & Q \\
11 & \nodata & 66 & \nodata & \nodata & \nodata & \nodata & 260.1375430 & 65.7552110 & 0.001$^{a}$ & 40.31 & 10.96 & 2.27 & 6 & 0.49 & Q \\
16 & \nodata & 162 & \nodata & \nodata & \nodata & \nodata & 260.1518860 & 65.8707960 & 0.876$^{a}$ & 0.48 & 2.20 & 0.64 & 5 & 0.44 & Q \\
23 & 194 & \nodata & \nodata & \nodata & \nodata & \nodata & 260.1714480 & 65.9675900 & 2.188 & \nodata & 9.08 & 1.72 & 5 & 0.30 & G \\
24 & \nodata & 221 & \nodata & \nodata & \nodata & \nodata & 260.1745000 & 65.6967160 & 4.492$^{a}$ & 1.04 & 12.52 & 12.82 & 5 & 0.61 & Q \\
25 & 209 & 250 & \nodata & \nodata & \nodata & \nodata & 260.1715390 & 65.7739260 & 0.172 & \nodata & 3.74 & 0.47 & 7 & 0.05 & G \\
& & & & & & & \multicolumn{1}{c}{$\cdots$} & & & & & & & & \\
227 & 49 & 15 & 3 & 19 & 38 & 218 & 260.6908870 & 65.7380750 & 1.713$^{a}$ & 0.45 & 28.58 & 1.54 & 15 & 0.85 & Q \\
229 & 69 & 50 & 44 & 32 & \nodata & \nodata & 260.6926570 & 65.8049470 & 1.610$^{a}$ & 1.59 & 9.65 & 1.91 & 7 & 0.42 & Q \\
230 & 167 & \nodata & \nodata & \nodata & \nodata & \nodata & 260.6931760 & 65.7526250 & 2.177$^{a}$ & 0.22 & 4.89 & 1.34 & 8 & 0.30 & Q \\
231 & \nodata & 206 & \nodata & \nodata & \nodata & \nodata & 260.6951290 & 65.5966950 & 4.682$^{a}$ & \nodata & 28.54 & 5.35 & 5 & 0.16 & G \\
233 & 35 & 29 & 112 & \nodata & 74 & \nodata & 260.6979060 & 65.7779390 & 2.781$^{d}$ & \nodata & 446.66 & 21.44 & 13 & 0.76 & Q \\
234 & 136 & \nodata & \nodata & \nodata & \nodata & 230 & 260.7008060 & 65.8243410 & 0.474 & \nodata & 84.52 & 9.43 & 7 & 0.20 & G \\
235 & 84 & 168 & \nodata & \nodata & \nodata & \nodata & 260.7017820 & 65.8127980 & 0.856$^{a}$ & \nodata & 9.69 & 0.49 & 16 & 0.61 & Q \\
237 & 193 & 21 & 76 & \nodata & 21 & \nodata & 260.7043150 & 65.8897710 & 0.001 & \nodata & 41.02 & 1.93 & 15 & 0.01 & S \\
& & & & & & & \multicolumn{1}{c}{$\cdots$} & & & & & & & & \\
426 & \nodata & 154 & \nodata & \nodata & \nodata & \nodata & 261.1939700 & 65.7260060 & 4.400$^{a}$ & 0.60 & 9.48 & 1.06 & 5 & 0.13 & G \\
427 & \nodata & 196 & \nodata & \nodata & \nodata & \nodata & 261.2048650 & 65.7129970 & 4.568$^{a}$ & 1.14 & 16.13 & 2.36 & 5 & 0.09 & G \\
428 & \nodata & 219 & \nodata & \nodata & \nodata & \nodata & 261.2092900 & 65.7330470 & 1.556$^{a}$ & 6.15 & 2.73 & 0.44 & 5 & 0.81 & Q \\
429 & 224 & \nodata & \nodata & \nodata & \nodata & \nodata & 261.2097470 & 65.9133070 & 0.826 & \nodata & 3.17 & 0.44 & 5 & 0.22 & G \\
433 & \nodata & 18 & 106 & \nodata & 64 & \nodata & 261.2291870 & 65.7214810 & 4.219$^{a}$ & 5.26 & 5.51 & 3.23 & 5 & 0.58 & Q \\
435 & \nodata & 274 & \nodata & \nodata & \nodata & \nodata & 261.2382200 & 65.9517820 & 0.590$^{e}$ & \nodata & 11.64 & 3.62 & 5 & 0.59 & Q \\
442 & 174 & \nodata & \nodata & \nodata & \nodata & \nodata & 261.2700500 & 65.8646700 & 4.455$^{a}$ & 6.22 & 5.98 & 4.88 & 5 & 0.22 & G \\
447 & \nodata & 119 & \nodata & \nodata & \nodata & \nodata & 261.3178710 & 65.8920670 & 4.255$^{a}$ & 12.52 & 11.82 & 12.71 & 5 & 0.84 & Q \\
\enddata
\tablecomments{This table is published in its entirety in machine-readable format. A portion is shown here for guidance regarding its form and content. XMM-IDs are the same as those of Silver et al., ApJ, submitted (2026). Superscripts on $z$ denote: $^{a}$ photo-$z$ solution from the maximally likely CIGALE $P(z)$; $^{b}$ new DESI spectroscopic redshift; $^{c}$ new DECaLS spectroscopic redshift; $^{d}$ new NIRISS spectroscopic redshift; $^{e}$ new binospec2025 spectroscopic redshift.}
\end{deluxetable*}

\end{document}